\documentclass[reprint,amsmath,amssymb,aps,nofootinbib,twocolumns]{revtex4-2}
\usepackage{lipsum} % use as '\lipsum[2-4]' for dummy text
\usepackage{xcolor} % for \textcolor
\usepackage[T1]{fontenc}

\usepackage[letterpaper,top=1.45cm,bottom=1.65cm,left=2cm,right=2cm]{geometry}
\usepackage{graphicx}% Include figure files
\usepackage{dcolumn}% Align table columns on decimal point
\usepackage{bm}% bold math
\usepackage{hyperref}% add hypertext capabilities
\usepackage{multirow}

\graphicspath{{images/}} % images' folder
\usepackage{lipsum} % use as '\lipsum[2-4]' for dummy text
\usepackage{xcolor} % for \textcolor

\usepackage[normalem]{ulem} % \sout command to strikethrough text

\begin{document}

\title{The endpoint of the Gregory-Laflamme instability of black strings revisited}

\author{Pau Figueras}%
 \email{p.figueras@qmul.ac.uk}
\affiliation{%
 School of Mathematical Sciences, Queen Mary University of London\\
 Mile End Road, London, E1 4NS, United Kingdom
}%

\author{Tiago França}
 \email{t.e.franca@qmul.ac.uk}
\affiliation{%
 School of Mathematical Sciences, Queen Mary University of London\\
 Mile End Road, London, E1 4NS, United Kingdom
}%

\author{Chenxia Gu}
 \email{chenxia.gu@qmul.ac.uk}
\affiliation{%
 School of Mathematical Sciences, Queen Mary University of London\\
 Mile End Road, London, E1 4NS, United Kingdom
}%

\author{Tomas Andrade}%
\email{tandrade@icc.ub.edu}
\affiliation{%
 Departament de F\'isica Qu\`antica i Astrof\'isica, Institut de Ci\`encies del Cosmos, Universitat de Barcelona,\\
Mart\'i i Franqu\`es 1, E-08028 Barcelona, Spain
}%

\begin{abstract}
    We reproduce and extend the previous studies of Lehner and Pretorius of the endpoint of the Gregory-Laflamme instability of black strings in five space-time dimensions. We consider unstable black strings of fixed thickness and  different lengths, and in all cases we confirm that at the intermediate stages of the evolution the horizon can be interpreted as a quasistationary self-similar sequence of black strings connecting spherical black holes on different scales. However, we do not find any evidence for a global timescale relating subsequent generations. The endpoint of the instability is the pinch off of the horizon in finite asymptotic time, thus confirming the violation of the weak cosmic censorship conjecture around black string spacetimes. 
\end{abstract}

\maketitle

%%%%%%%%%%%%%%%%%%%%%%%
\section{Introduction}
%%%%%%%%%%%%%%%%%%%%%%%

General Relativity (GR) is the currently accepted classical theory of gravity. Quite remarkably, so far it has successfully passed all experimental tests, explaining gravitational phenomena on an incredibly wide range of scales, from Solar system scales to cosmology. Furthermore, the detections of gravitational waves by the LIGO/Virgo/KAGRA collaboration \cite{LIGOScientific:2016aoc,LIGOScientific:2016sjg} produced in mergers of compact objects have allowed for new  tests of Einstein's theory in the strong field regime; such tests should lead to new insights into fundamental aspects of the theory.  

Black holes play a very important role in our understanding of GR, and gravity in general, due to their simplicity, which makes them tractable, and the fact that they capture key aspects of the theory. One of the distinguishing features of black holes is the presence of singularities in their interior, where the description provided by GR breaks down. Penrose's famous singularity theorem \cite{Penrose:1964wq} establishes that singularities in GR can occur more generally, as one should expect in a non-linear theory, and they need not be hidden inside black holes. The occurrence of singularities limits the predictivity of GR as a classical theory of gravity; to ensure that GR retains its predictive power in the presence of certain singularities,  Penrose conjectured that the latter should \textit{generically} be cloaked by horizons. This is the Weak Cosmic Censorship Conjecture (WCCC) \cite{Penrose:1969pc} (see \cite{Geroch:1978ub,Geroch:1979uc,Christodoulou_1999} for a modern and mathematically precise formulation), and there are no known counter-examples in four-dimensional astrophysical settings.\footnote{Choptuik's critical collapse \cite{Choptuik:1992jv} is non-generic. In fact, Christodoulou proved the WCCC for the Einstein-scalar field model in four dimensional asymptotically flat spacetimes in spherical symmetry \cite{Christodoulou:1999aa}.} Proving or disproving this conjecture remains one of the most important open problems in mathematical relativity (see \cite{Christodoulou:2008nj} for recent progress).

While the Kerr black hole is believed to be stable \cite{Whiting:1988vc,Dafermos:2010hb,Dafermos:2014cua,Dafermos:2021cbw,Klainerman:2021qzy}, and hence its relevance in astrophysics, higher dimensional vacuum black holes exhibit much richer dynamics. Almost 30 years ago, Gregory and Laflamme (GL) discovered that black strings (and black $p$-branes in general) can be unstable to long wavelength perturbations that break the translational symmetry along the compact extra dimensions \cite{Gregory:1993vy}. It is fair to say that the study of the GL instability and its possible endpoints together with the discovery of the asymptotically flat black ring in five dimensions \cite{Emparan:2001wn} were largely responsible for the intrinsic interest in understanding the physics of higher dimensional black holes regardless of string theory (see \cite{Emparan:2008eg} for a review). From the extensive work done in this area over the years it has become clear that the GL instability is very general and it basically affects any higher dimensional black hole which is sufficiently far from extremality whenever the horizon geometry is characterized by widely separated length scales. The latter happens for instance for rapidly spinning asymptotically flat black holes \cite{Emparan:2003sy,Dias:2009iu,Dias:2010eu,Dias:2014eua} and black rings \cite{Santos:2015iua}, or in anti-de Sitter black holes \cite{Hubeny:2002xn,Hirayama:2001bi}. Therefore, the GL instability can teach us very general aspects of the physics and the possible phases of black holes and their dynamics in a wide variety of settings.

The endpoint of the GL instability of black strings was finally spelled out in a famous paper by Lehner and Pretorius \cite{Lehner:2010pn} (see \cite{Lehner:2011wc} for a reivew), who used numerical relativity techniques to solve the Einstein equations numerically in the highly dynamical and fully non-linear regime. They found that the horizon evolves into a sequence of spherical black holes joined by string segments; these string segments are themselves GL unstable, triggering a cascade of instabilities that give rise to new generations of black holes and black strings on ever smaller scales, eventually leading to the pinch off of horizon somewhere along the string. In particular, \cite{Lehner:2010pn} argue that the dynamics of the GL instability is (globally) self-similar.\footnote{This should not be confused with local self-similarity near the singularity, which would manifest itself as a scaling solution describing the approach to the singularity.} Using this observation they estimate that the pinch off time (as measured by asymptotic observers) is given by a geometric series and hence finite. Since horizons cannot bifurcate in a smooth manner (see e.g., Proposition 9.2.5 in \cite{Hawking:1973uf}), \cite{Lehner:2010pn} conclude that a naked singularity must form at the pinch off. Indeed, the simulations of \cite{Lehner:2010pn} show that the spacetime curvature invariants at the horizon of the black string blow up as the system approaches the pinch off. Furthermore, since no fine tuning is required, one concludes that the endpoint of the GL instability constitutes a violation of the WCCC in higher dimensional asymptotically Kaluza-Klein (KK) spaces. Following the original work of \cite{Lehner:2010pn}, further studies of certain higher dimensional asymptotically flat rotating black holes and black rings that also suffer from the GL instability  have provided further support for this picture in finite number of spacetime dimensions \cite{Figueras:2015hkb,Figueras:2017zwa,Bantilan:2019bvf,Andrade:2020dgc} and in the large $D$ limit of GR \cite{Andrade:2018yqu,Andrade:2019edf,Andrade:2020ilm,Emparan:2021ewh}.

The seminal paper of \cite{Lehner:2010pn} is more than ten years old and, as important as it is for the current understanding of the WCCC and its potential violations, it has not been independently reproduced in the literature. Therefore, the first goal of the present article is to reproduce the main results of \cite{Lehner:2010pn}, using completely independent methods and a different code: While \cite{Lehner:2010pn} solve the Einstein equations using harmonic coordinates and excision with the \texttt{PAMR/AMRD} libraries,\footnote{\texttt{http://laplace.physics.ubc.ca/Group/Software.html}.} here we solve the Einstein equations using the CCZ4 formulation \cite{Alic:2011gg,Alic:2013xsa} in singularity avoiding coordinates and the \texttt{GRChombo} code \cite{Clough:2015sqa,Andrade:2021rbd}.  The second goal of this paper is to extend the results of \cite{Lehner:2010pn} by evolving the system closer to the singularity than ever before and by considering black strings of different lengths to obtain a more general picture of the evolution and the endpoint of the GL instability of black strings. While we reproduce the main result of \cite{Lehner:2010pn}, namely that the GL unstable black string evolves into a sequence of ever thinner strings connecting black holes leading to a pinch off in finite asymptotic time, we do not find any evidence of a global timescale relating subsequent generations. This implies that the pinch off time is not given by a geometric series; instead, the local dynamics on the string segments plays an important role beyond the third generation, leading to a faster approach to the singularity. The distinct role that the third generation plays here is due to our choice of initial data (as well as in \cite{Lehner:2010pn}). Furthermore, our simulations indicate that the  dynamics near the singularity seems to be independent of macroscopic details of the string, suggesting the existence of a universal local solution that controls the pinch off, as in certain fluids \cite{Eggers:1993aa,Eggers:1997aa}.

The rest of this article is organised as follows: In Section \ref{sec:methods} we describe in detail the numerical methods that we have used and our choice of initial conditions. Section \ref{sec:results} contains the main results of the article. In Section \ref{sec:evol_AH_area} we describe the evolution of the apparent horizon area; Section \ref{subsec:dynamics_AH} studies the dynamics of the apparent horizon and Section \ref{sec:singularity} contains the details of the approach to the singularity. In Section \ref{sec:discussion} we summarize our main results and outline directions for future research. Convergence tests are presented in Appendix \ref{app:convergence}. 

In this article we use the following conventions: $G=c=1$. Greek letters $\mu,\nu,\ldots$ denote spacetime indices while Latin letters $i,j,\ldots$ denote indices on the spatial hypersurfaces. 

%%%%%%%%%%%%%%%%%%%%%%%
\section{Numerical Methods}
\label{sec:methods}
%%%%%%%%%%%%%%%%%%%%%%%

%%%%%%%%%%%%%%%%%%%%%%%
\subsection{Evolution}
%%%%%%%%%%%%%%%%%%%%%%%

We solve the Einstein vacuum equations in 4+1 dimensions in the CCZ4 formulation \cite{Alic:2011gg,Alic:2013xsa} using the \texttt{GRChombo} code \cite{Clough:2015sqa,Andrade:2021rbd}. We use Cartesian coordinates and impose SO(3) symmetry along the Minkowski directions at the level of the equations of motion using the modified cartoon method \cite{Pretorius:2004jg,Shibata:2010wz,Cook:2016soy}, thus reducing the effective dimensionality of the problem to 2+1. The dimensionally reduced equations of motion in the BSSN formulation can be found in \cite{Cook:2016soy}; the generalization to CCZ4 is straightforward. 

To stably simulate black string spacetimes, we redefine the constraint damping parameter $\kappa_1\to \kappa_1/\alpha$ as in \cite{Alic:2013xsa}, where $\alpha$ is the lapse function. In the results reported in Section \ref{sec:results}, we used $\kappa_1=0.37$ and $\kappa_2=-0.8$. We use $6^\text{th}$ order finite differences to discretise the spatial derivatives and a standard RK4 time integrator to step forward in time. Since the overall convergence order cannot be higher than four, we use  $6^\text{th}$ order Kreiss-Oliger dissipation. In Appendix \ref{app:convergence} we show that the order of convergence that we achieve is roughly three, as expected in a typical AMR code as \texttt{GRChombo}.  As in similar settings \cite{Figueras:2015hkb,Figueras:2017zwa,Bantilan:2019bvf}, to control the gradients near the coordinate singularity present in the computational domain inside black holes, we add diffusion terms well inside the apparent horizon (AH) to the right hand side of the equations of motion for those variables that appear with second order spatial derivatives. We place the outer boundary along the Minkowski directions at $L_\text{outer}=256r_0$, where $r_0$ is the mass parameter of the black string, see equation \eqref{eq:Schw_GP} in Section \ref{sec:init_data}. At the outer boundary $x=L_\text{outer}$ we impose either Sommerfeld or periodic boundary conditions, while the direction along the string, $z$, is periodic with period $L$.\footnote{Since $L_\text{outer}$ is not in causal contact with the black string for the entire duration of our simulations (see Section \ref{sec:results}), the particular choice of boundary conditions at $x=L_\text{outer}$ makes no difference in practice.}  The grid spacing in the coarsest level typically is $dx=0.25\,r_0$ and we add another 12 levels of refinement (so 13 levels in total) with a refinement ratio of 2:1. 

We evolve the lapse $\alpha$ with the standard 1+log slicing condition, 
\begin{equation}
    (\partial_t-\beta^i\partial_i)\alpha = c_\alpha\,\alpha\,(K-2\,\Theta)\,,
    \label{eq:lapse_eq}
\end{equation}
with $c_\alpha = 1.3$. Here $K$ is the trace of the extrinsic curvature of the spatial slices and $\Theta$ is another of the CCZ4 evolution variables. We evolve the shift vector $\beta^i$ with the integrated Gamma-driver,
\begin{equation}
    (\partial_t-\beta^j\partial_j)\beta^i=c_\beta\,\hat \Gamma^i-\eta\,\beta^i\,,
    \label{eq:gamma_driver}
\end{equation}
where $\hat\Gamma^i$ is the usual CCZ4 evolution variable and $c_\beta=0.6$; these choices of gauge parameters have proven to work well in numerical simulations of higher dimensional black hole spacetimes \cite{Shibata:2009ad,Shibata:2010wz,Figueras:2015hkb,Figueras:2017zwa,Bantilan:2019bvf}.\footnote{Note that the values of the gauge parameters $c_\alpha$ and $c_\beta$ that we use in \eqref{eq:lapse_eq} and \eqref{eq:gamma_driver} differ from the typical values used in black holes binary mergers in astrophysical scenarios, which  are $c_\alpha=2$ and $c_\beta=0.75$.}  Notice that unlike \cite{Figueras:2015hkb,Figueras:2017zwa,Bantilan:2019bvf}, we have not included an extra term in \eqref{eq:gamma_driver} corresponding to the contracted Christoffel symbols of the (conformally rescaled) initial spatial metric. The reason is that such term vanishes for our choice of initial conditions, see Section \ref{sec:init_data}.

\subsection{Initial data}
\label{sec:init_data}
We start with an unperturbed 5$D$ black string written in Gullstrand-Painlev\'e coordinates \cite{Painleve,Gullstrand},
\begin{equation}
ds^2= -\left(1-\frac{r_0}{r}\right)dt^2+2\sqrt{\frac{r_0}{r}}dt\,dr+dr^2+r^2d\Omega_{(2)}^2 + dz^2\,,
\label{eq:Schw_GP}
\end{equation}
where $r_0$ is the usual Schwarzschild mass parameter, $z\sim z+L$ is the KK compact direction and $d\Omega_{(2)}^2$ is the standard metric on the unit round two-sphere. The ADM mass of \eqref{eq:Schw_GP} is 
\begin{equation}
M = \tfrac{1}{2}\,L\,r_0\,.
\label{eq:ADM_mass}
\end{equation}

The advantage of using these coordinates is that they are horizon-penetrating while the metric on the spatial sections is flat. The latter makes it particularly convenient for constructing perturbed initial data that minimizes the amount of initial constraint violations, as we explain below. We regularize the physical singularity present in the initial data slice using the ``turduckening" approach \cite{Brown:2007pg,Brown:2008sb} and cutting off by hand the range of the radial coordinate. In terms of the radial cartoon coordinate $x$, if $x<\varepsilon$, then we evaluate the initial data quantities derived from \eqref{eq:Schw_GP} at $x=\varepsilon$; for the results presented below, we typically use $\varepsilon=0.1\,r_0$.\footnote{We should emphasize that we only employ this regularization procedure at the level of the initial data; at the later stages in the evolution, the $x$ coordinate takes values in $0<x<x_\text{max}$.}  

From the initial data \eqref{eq:Schw_GP}, we read off the $4+1$ quantities, noting that in Cartesian coordinates $\det\gamma =1$ and hence the unperturbed conformal factor satisfies $\chi \equiv (\det\gamma )^{-\frac{1}{4}}=1$. Furthermore,  the Christoffel symbols associated to the spatial conformal metric $\tilde\gamma_{ij}$ trivially vanish. We introduce a constraint violating perturbation on the conformal factor $\chi$ that triggers the GL instability:
\begin{equation}
\chi = \textstyle 1+\epsilon\,\sin\left(\frac{2\pi n z}{L}\right)\,e^{-\left(\frac{x}{r_0}-\frac{r_0}{x}\right)^2}\,,
\label{eq:chi_p}
\end{equation} 
where $\epsilon$ is the amplitude of the perturbation, $n\in \mathbb{N}$ selects the GL harmonic to be excited and the exponential factor in \eqref{eq:chi_p} ensures that the perturbation is localized near the horizon. Therefore, our perturbation \eqref{eq:chi_p} does not change the ADM mass \eqref{eq:ADM_mass} of the spacetime. For the results reported in Section \ref{sec:results}, we chose $r_0=1$, $\epsilon=0.01$  and $n=1$. We keep track of the constraint violations introduced by our perturbation and we verify that for $\epsilon\lesssim 0.01$ they are exponentially suppressed by the damping terms in the CCZ4 equations on a timescale that is much faster than any other timescale in the problem. The remaining $4+1$ quantities are left unperturbed; for the initial lapse $\alpha$ and shift vector $\beta^i$,  we  set  $\alpha=1$ and $\beta^i=0$.\footnote{This choice of lapse and shift implies that initially there is a period of strong gauge dynamics superposed with the physical evolution of the GL instability. This period of gauge adjustment typically lasts $t/r_0\sim 10$, whilst with our perturbation the onset of the GL instability occurs at $t/r_0\sim 70-100$ or even later, see Fig. \ref{fig:AH_area}.}

%%%%%%%%%%%%%%%%%%%%%%%
\subsection{Grid hierarchy and AMR}
%%%%%%%%%%%%%%%%%%%%%%%

The location along the string where the first generation forms is sensitive to the initial conditions, see Section \ref{sec:results}. Beyond this point and for initial data with zero total momentum, as in our case and in \cite{Lehner:2010pn}, the evolution should respect the $\mathbb Z_2$ reflection symmetry about the centre of the first generation string segment. The reason is that the $n=1$ GL harmonic, which is the one that governs the subsequent universal evolution of the strings with the thicknesses that we have considered, has this symmetry.  
However, truncation errors due to the asymmetry of the various levels of refinement that are automatically added as the evolution of the instability unfolds can break this physical symmetry of the problem. This effect is visible from the third generation onwards in the animation of the evolution of the instability produced by \cite{Lehner:2010pn} and in Fig. 2 of this reference. In this subsection we provide details of the grid hierarchy that we used in our simulations that ensures that our numerical method respects this $\mathbb Z_2$ symmetry; we emphasise that we did not explicitly enforce this symmetry in our simulations, so the fact that it is respected is a sign that the truncation errors do not interfere with the physics. The reader interested in the results of our simulations can safely skip the rest of this subsection. 

In similar problems that some of us addressed in the past using \texttt{GRChombo}, e.g., \cite{Figueras:2015hkb,Figueras:2017zwa,Bantilan:2019bvf,Andrade:2020dgc}, a refinement criterion based on the gradients of the conformal factor $\chi$ as estimates of the local numerical error was used. The reason is that in the moving punctures gauge (i.e., $1+\log$ slicing for the lapse plus Gamma driver for the shift), the contours of the conformal factor $\chi$ track the AH quite accurately. Therefore, such a criterion leads to hierarchy of refinement levels that approximately follows the shape of the AH, thus potentially optimizing the computational cost of the simulations. However, in the problems considered in \cite{Figueras:2015hkb,Figueras:2017zwa,Bantilan:2019bvf,Andrade:2020dgc}, one did not have to worry about any underlying symmetries that the numerical method ought to respect, so this refinement criterion turned out to be very convenient and efficient. Unfortunately, in the present case of the black string, we were not able to find an efficient way to dynamically ensure that the grid hierarchy that resulted from a refinement criterion based on the gradients of $\chi$ respected the $\mathbb Z_2$ symmetry of the problem.

\begin{figure}[t]
\centering
\includegraphics[width=0.48\textwidth]{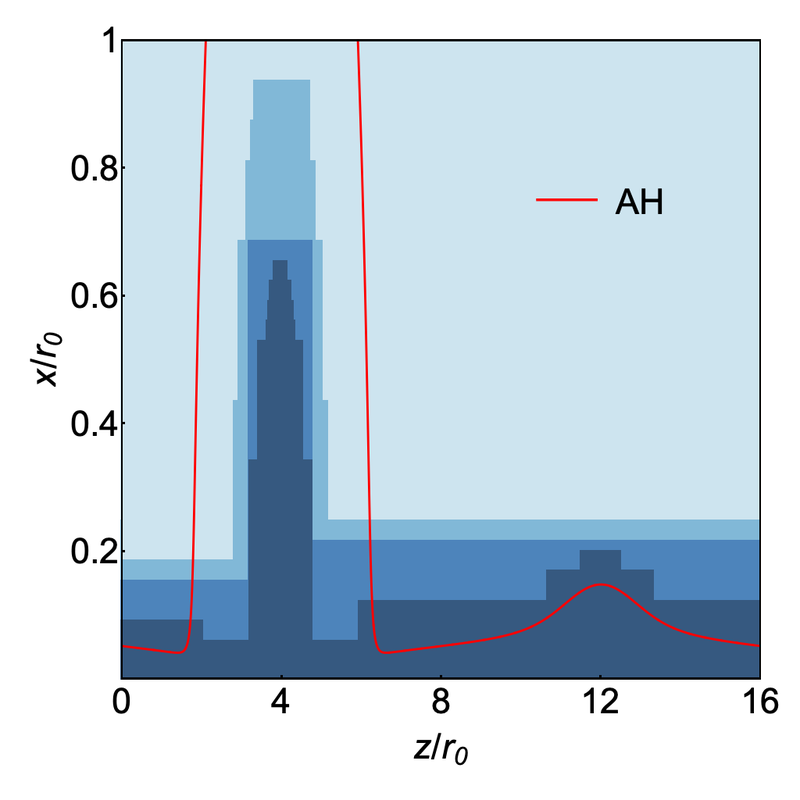}
\includegraphics[width=0.48\textwidth]{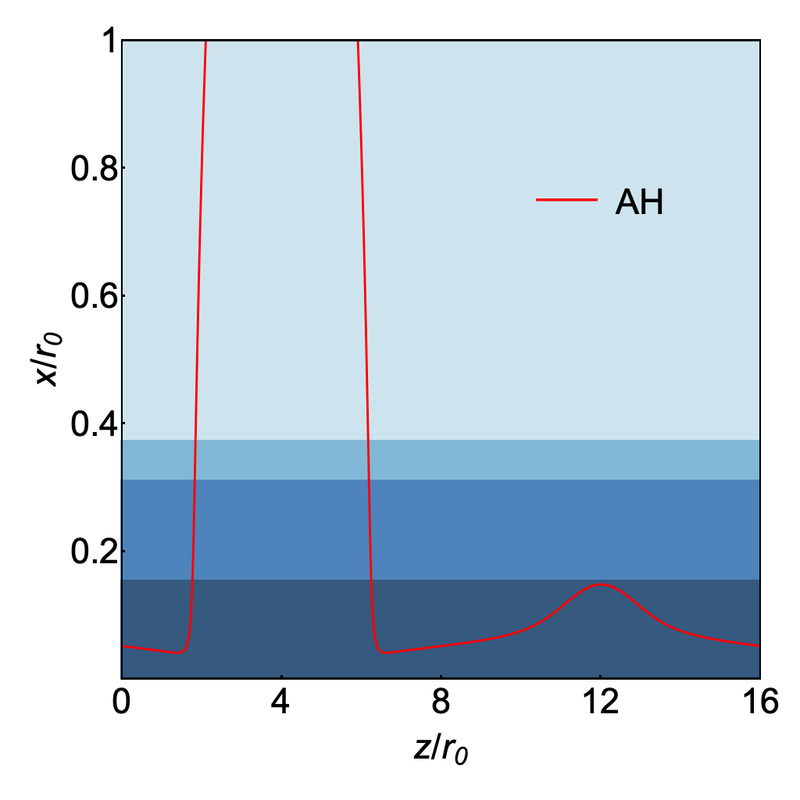}
\caption{Grid hierarchy obtained with a dynamical refinement criterion based on the gradients of the evolution variable $\chi$ (top) versus a rectangular grid hierarchy determined by the resolution at the AH (bottom). The actual location of the AH is depicted in red and the color gradient corresponds to the different refinement levels, with darker blue depicting finer (i.e., higher resolution) levels. In this example we considered a black string with $r_0=1$ and $L=16$.}
\label{fig:levels}
\end{figure}

In the top panel in Fig. \ref{fig:levels} we show a typical example of the hierarchy of innermost refinement levels that result from using this tagging criterion. As this figure shows, the levels are not symmetric. As consequence, the high frequency noise that the AMR algorithm produces at the level boundaries can lead to an accumlation of errors that can result in late time unphysical effects, typically beyond the third generation, unless very high resolution is used.   Another issue with a $|\partial\chi|$-based refinement criterion in the black string problem is that the finest level covers a significant portion of the interior of the big blob, which seems wasteful, see the top panel of Fig. \ref{fig:levels}.  
We tried to minimize the size of this region covered by the finest levels, but the resulting AMR hierarchy led to large truncation errors at the level boundaries that were hard to control in practice.  Another potential issue with a dynamical tagging criterion such as the $|\partial\chi|$-based one is that it inevitably leads to significant AMR level dynamics, in particular when new generations appear since the variable $\chi$ changes very rapidly in some regions, resulting in frequent regridding and hence higher levels of numerical errors. We tried other dynamical refinement criteria based on the trace of the extrinsic curvature $K$ or the CCZ4 variable $\hat \Gamma^i$ but the results were qualitatively the same.

To ensure that the truncation errors do not interfere with the dynamics of the unstable black strings, we enforce that all refinement levels are rectangular-shaped, see the bottom panel in Fig. \ref{fig:levels}. This shape of the refinement levels ensures that our numerical method respects the $\mathbb Z_2$ symmetry of the system but does not enforce it, and the  computational cost of the simulations is comparable to that of a $|\partial\chi|$-based refinement criterion.   The thinnest parts of the AH of the black string determine the necessary resolution of the finest level, which in turn essentially  determines the computational cost of the simulation. In practice, we require that the AH is covered by at least 40 grid points, but typically we had 60 or more grid points covering the AH; if there are not enough points in the finest level, a new level is added.\footnote{Note that because of the reflection symmetry about $x=0$, our criterion ensures that the whole of the AH is covered by at least 80 points.} In addition, we monitor the constraint violations throughout the simulation and verify that they remain under control. With rectangular refinement levels, some of the not-so-thin parts of the AH are overly resolved (see the bottom panel in Fig. \ref{fig:levels}) but,  on the other hand, the finest refinement level only covers a small portion of the interior of the big blob. Furthermore, with rectangular levels, once a new level is added, we do not regrid anymore and hence its shape is fixed once and for all, thus minimizing the number of interpolation/extrapolation operations at the level boundaries. This helps to control the errors and speeds up the simulation. 

We note that the computational cost of the simulation increases exponentially as the singularity is approached if we insist in requiring that the AH is covered by at least 40 points, as we do. With more levels added to the grid hierarchy, the size of the checkpoint files likewise increases exponentially, which can eventually pose a storage problem.  Therefore, even with AMR, with finite resources one cannot possibly reach the pinch off. Summarizing, the computational and storage costs of the simulations of the GL instability of black strings eventually become prohibitive; for the results presented in this article, the resources required towards the end of the simulations exceed those of a typical equal mass non-spinning black hole binary merger, which is a $3+1$ problem.

%%%%%%%%%%%%%%%%%%%%%%%
\subsection{Apparent horizon}
%%%%%%%%%%%%%%%%%%%%%%%

We assume that the AH is given by the level set $F=0$ of the function
\begin{equation}
    F(x,z)= x - h(z)\,.
    \label{eq:ah_eq}
\end{equation}
Imposing that the outgoing null rays have vanishing expansion on this surface gives a non-linear second order ODE for $h(z)$. We solve this equation using the \texttt{PETSc} libraries \cite{petsc-efficient}. Typically we do so at every full time step of the coarsest level and we use the previous solution as the initial guess. 

The ansatz \eqref{eq:ah_eq} for the shape of the AH is the same as in \cite{Choptuik:2003qd,Lehner:2010pn}; even though the AH is slice-dependant and our gauge is different from the ones used in these references, the embedding diagrams of the AH that we have obtained (see below and Fig. \ref{fig:AH_embedding}) look indistinguishable from those in \cite{Choptuik:2003qd,Lehner:2010pn}. This is not unrelated to the fact that, as shown in \cite{Choptuik:2003qd}, for most of the evolution the AH is a very good approximation to the event horizon.

To get some intuition about the shape of the AH during the evolution, we produce embedding diagrams of the geometry into $\mathbb R^4$ following \cite{Choptuik:2003qd,Lehner:2010pn}:
\begin{equation}
    ds^2 = dR^2+R^2\,d\Omega_{(2)}^2+dZ^2
\end{equation}
where $Z$ is periodic.  The embedding coordinate $R$ is defined as the areal radius of the horizon $S^2$; in our working coordinates this is, 
\begin{equation}
    R(z)=\sqrt{g_{ww}}\,h(z)\,,
\end{equation}
where $w$ denotes any of the Cartesian cartoon directions along the $S^2$. Then, the other embedding coordinate $Z$ is uniquely defined by demanding that the proper length of the AH along the compact direction $z$ is the same as the Euclidean length of the curve $R(Z)$ in the embedding diagram. This gives,
\begin{equation}
    Z(z)=\int^z_0 d\bar z\,\frac{\sqrt{g_{zz} + 2\,g_{zx}\,h'(\bar z) + g_{xx}\,h'(\bar z)^2}}{\sqrt{{h'}(\bar z)^2+1}}\,.
\end{equation}

Finally, we monitor the spacetime Kretschmann invariant evaluated on the AH:
\begin{equation}
    \bar{\mathcal{K}} = \frac{1}{12}\,R_{abcd}\,R^{abcd}\,R^4\,,
    \label{eq:kretschmann}
\end{equation}
where the normalization has been chosen so that $\bar{\mathcal{K}}=1$ at the horizon of a stationary uniform black string, and $\bar{\mathcal{K}}=6$ at the horizon of an asymptotically flat 5$D$ Schwarzschild black hole.

%%%%%%%%%%%%%%%%%%%%%%%
\section{Results}
\label{sec:results}
%%%%%%%%%%%%%%%%%%%%%%%

In this section we present our results. In our simulations, we have fixed the mass parameter of the parent 4$D$ Schwarzschild black hole to $r_0=1$ (see eq. \eqref{eq:Schw_GP}) and varied the asymptotic length of the compact circle $L$, and hence the total mass. We discuss the results for the different values of $L$ that we have considered simultaneously in order to present the general picture of the dynamics of the GL instability. As we shall explain below, the computational cost differs for different $L$'s and we were not able to get equally close to the singularity for all values of $L$ due to our limited computational resources. We aimed at pushing the $L=10$ case as much as we could since this is the case that can be directly compared with the results of \cite{Lehner:2010pn}.\footnote{Note that these references present their results in units of the initial $4D$ Schwarzschild mass $M_{\text{Schw}}=r_0/2=1$, so their results and ours are related by an overall rescaling by a factor of 2.} The equivalent case to $L=10$ with $r_0=1$ was chosen in \cite{Choptuik:2003qd,Lehner:2010pn,Lehner:2011wc} because it leads to approximately the fastest growth rate for the shortest wavelength instability. In practice we have found that with our perturbations \eqref{eq:chi_p} and the values of $L$ that we have considered,  the time that it takes for the dynamics to enter the non-linear regime is comparable in all cases. The $L=8$ case, which corresponds to a fatter string,  takes a bit longer to enter the non-linear regime, as expected.

Just as \cite{Lehner:2010pn,Lehner:2011wc}, we observe that the GL instability of a uniform black string evolves into a dynamical black hole that can be described as a quasistationary sequence of spherical black holes connected by thin (and unstable) black strings on different scales. The time of formation of the first generation is sensitive to the choice of initial conditions but the subsequent evolution is universal. Therefore, for the rest of this section, we will focus on describing the dynamics of the unstable black strings and the evolution of the AH after the first generation has formed. An animation of the evolution of the AH for the $r_0=1$ and $L=10$ black string can be found {\color{blue}\href{https://youtu.be/Mc-jzvn\_hto}{here}}; other animations corresponding to strings of different lengths can be found in the  {\color{blue}\href{https://youtube.com/playlist?list=PLSkfizpQDrcbUn2JNjkL0LKcy9k\_oGQ\_-}{\texttt{GRChombo} YouTube channel}}. 

\subsection{Evolution of the AH area}
\label{sec:evol_AH_area}

\begin{figure}[t]
\centering
\includegraphics[width=0.48\textwidth]{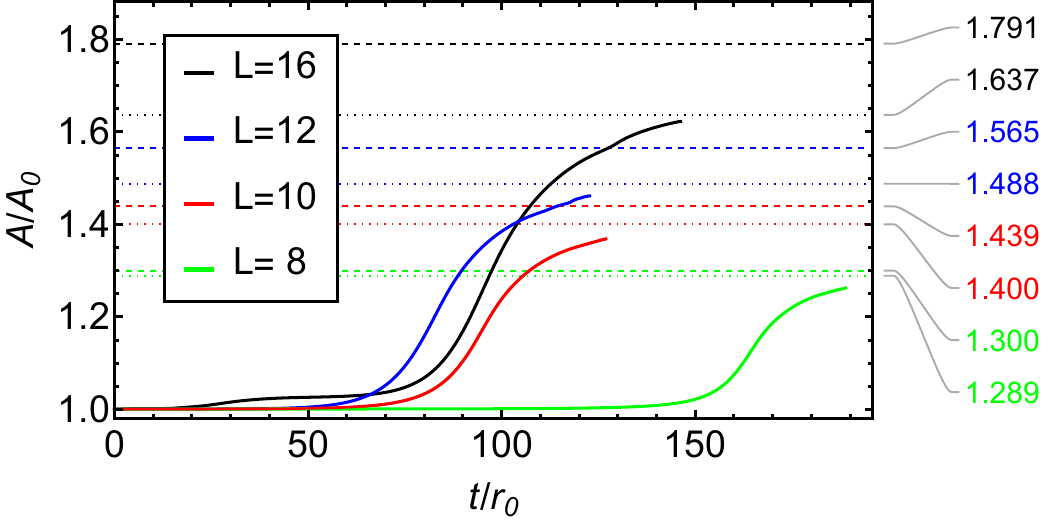}
\caption{Evolution of the AH area for unstable black strings in 5$D$ with 4$D$ mass parameter $r_0=1$ and different $L$'s. The dashed lines indicate the AH area of a slightly deformed  5$D$ KK black hole with the same mass as the unperturbed black string \cite{Harmark:2003yz}; the dotted lines correspond of the area of two black holes on a circle \cite{Dias:2007hg} with the same total mass as the unperturbed black string and the same radii as the first and second generation blobs measured at the end of the simulation.}
\label{fig:AH_area}
\end{figure}

In Fig. \ref{fig:AH_area} we display the evolution of the AH area normalized by the horizon area of the initial unperturbed black string of length $L$ and mass parameter $r_0=1$.  The evolution of the AH area in the $L=10$ case agrees very well with the results of \cite{Lehner:2010pn}, and it asymptotes to the area of an asymptotically flat 5$D$ Schwarzschild black hole with the same initial mass. For our choice of parameters, $A_{\text{Schw}_{5D}}/A_0=1.373$, while at the end of our simulation we have $A/A_0=1.369$. However, we believe that the closeness of these two numbers is just a coincidence and that the KK corrections to the horizon area should be taken into account. 

\begin{figure}[t]
\centering
\includegraphics[width=0.48\textwidth]{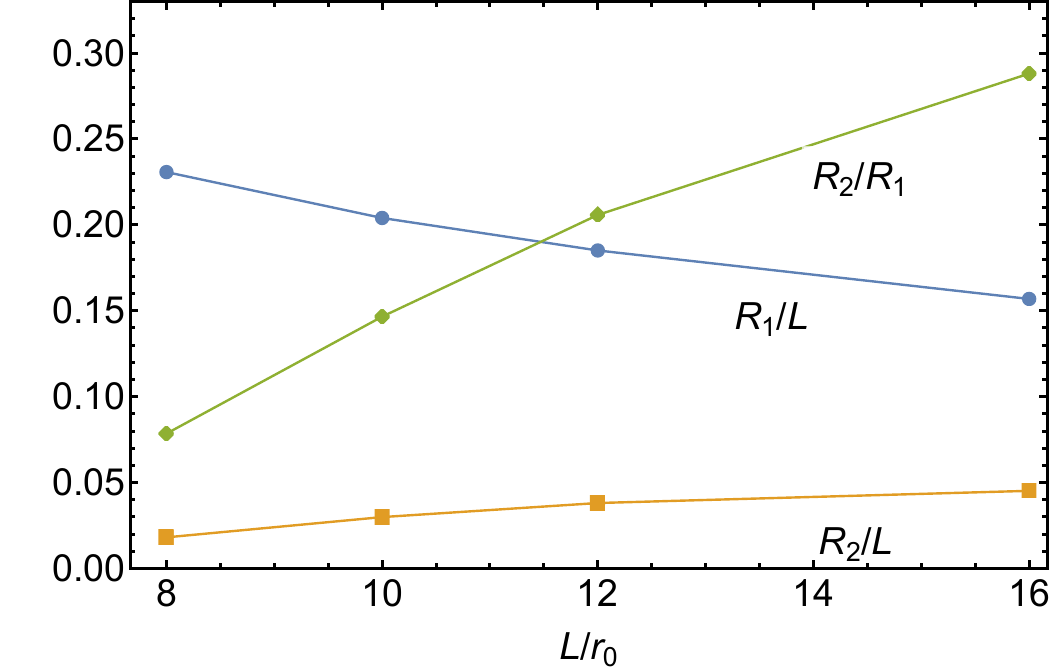}
\caption{Relative sizes of the first and second generation blobs as a function of the unperturbed string length $L$. The size of the second generation blob with respect to the first generation one increases with $L$. The lines have been added to guide the eye.}
\label{fig:relative_sizes}
\end{figure}

The dashed lines in Fig. \ref{fig:AH_area}  indicate the relative area of a slightly deformed single black hole on a circle with the same mass as the unperturbed black string with the same $L$. For the values of $L$ and masses that we have considered, the leading KK corrections to the thermondynamic quantities of a single 5$D$ caged black hole \cite{Harmark:2003yz} are between 6\% and $3\%$ of their respective asymptotically flat values, corresponding to the $L=8$ and $L=16$ cases respectively, so not negligible.\footnote{The area of an asymptotically flat 5$D$ Schwarzschild black hole with the same mass as the $L=8$ and $r_0=1$ uniform black string is $A/A_0=1.228$, which is less than the AH area at the end of our simulation. In this case, clearly the KK corrections cannot be ignored.}  If one takes these KK corrections into account, then the area of a single KK black hole with the same initial mass as the uniform black string with $r_0=1$ and $L=10$ is $A/A_0=1.439$, which is significantly larger than the AH area at the end of the simulation. We can see from Fig. \ref{fig:AH_area} that for larger values of $L$, i.e., $L=12$ and $L=16$, the final values of $A/A_0$ are even further away from the area of a single KK black hole with the same initial mass, while the difference is smaller for the $L=8$ case. The reason for this apparent disagreement is that we have ignored the contribution of the second generation bulge to the final area and the latter cannot be ignored, in particular for larger $L$, as it is larger than the size of the KK corrections. Indeed, the radius of the second generation bulge varies between 10\% and 30\% of the radius of the first generation bulge depending on $L$, see Fig. \ref{fig:relative_sizes}, and hence its contribution to the final area should also be taken into account. 

The dotted lines in Fig. \ref{fig:AH_area} show  that indeed the final total area compares better to the total area of two KK black holes with the same radii as the first and second generation bulges at the end of our simulations and same total mass as the initial unperturbed black string. This agreement gets better as $L$ increases. 
The reason  is that the size of the first generation bulge compared to the thickness of the parent string decreases as $L$ increases, see Fig. \ref{fig:relative_sizes}. In turn this leads to thicker first generation string segments and hence larger second generation bulges compared to the first generation blob. This is intuitive since as $L$ increases keeping the thickness of the string $r_0$ fixed, one would expect that the second GL harmonic becomes ``stronger'' until it would eventually dominate the dynamics for $L$ sufficiently large, leading to the formation of two blobs of equal sizes in the first generation.\footnote{We observe this effect at the third generation for the $L=8$ case, see Section \ref{subsec:dynamics_AH}.} Also, we see that the size of the third generation bulges compared to the second generation ones increases as $L$ decreases, see Fig. \ref{fig:AH_embedding}; hence, approximating the total area by that of only two KK black holes of the same  size as the first and second generation bulges is less accurate for smaller $L$.

The fact that the final AH area approaches the area of two static KK black holes on a circle with the same initial total mass indicates that the total amount of gravitational radiation emitted during the evolution of the GL instability up to this point is small.  It is likely that a strong burst of gravitational waves is emitted when the first generation forms; such a burst is known to occur in other systems that experience a GL instability and for which the gravitational waveforms have been computed \cite{Figueras:2017zwa,Andrade:2020dgc}. Beyond the formation of the first generation, the amount of gravitational wave emission decreases since the fast dynamics only takes place on ever decreasing scales and involving ever decreasing (local) masses, and it should be almost negligible at the the pinch off. Right after the pinch off one would expect that there is another period of strong gravitational wave emission since the dynamics of the system should be approximately that of a head-on collision between the first and second generation blobs. From the embedding diagrams in Fig. \ref{fig:AH_embedding}, we see that for larger $L$ the size of the second generation blob with respect to the first generation one increases, so the mass ratio $q$ of the two black holes in the collision would decrease. In the 5$D$ asymptotically flat case, \cite{Witek:2010az} computed the energy radiated in gravitational waves for head on collisions of black holes for different mass ratios and showed that it is less than 0.1\% of the initial total mass in all cases. Therefore, we expect that the total energy radiated via gravitational wave emission during the GL instability of black strings, including the pinch off and final state, should also be small. We hope to report on this in future work.

%%%%%%%%%%%%%%%%%%%%%%%
\subsection{Dynamics and geometry of the AH}
\label{subsec:dynamics_AH}
%%%%%%%%%%%%%%%%%%%%%%%

\begin{figure}[t]
\centering
\includegraphics[width=0.48\textwidth]{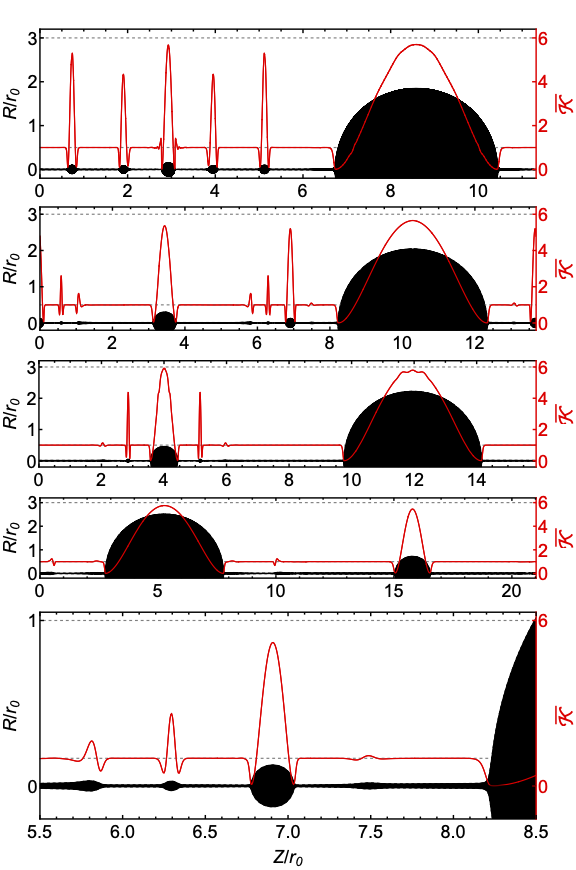}
\caption{\textit{Top}: Embeddings of the geometry of the AH at the last stage of our simulations for $L=8,\,10,\,12,\,16$, from top to bottom; the corresponding length along the $Z$ direction  is $11.32, 13.69, 15.90$ and $21.03$ in units of $r_0$, respectively. Note that we have rescaled the height of the plots so that all of them have the same width while maintaining the proportions of the embeddings. Superposed to the AH, we have plotted (in red) the induced Kretschmann invariant on the AH, suitably normalized \eqref{eq:kretschmann}.
\textit{Bottom}: zoom in of the thinnest part of the string in the $L=10$ case. The third and fourth generations are clearly visible. The latter is still rapidly evolving.}
\label{fig:AH_embedding}
\end{figure}

In Fig. \ref{fig:AH_embedding} we display the embeddings of the AH geometry at the last stage of our simulations for the various $L$'s that we have explored. In red we have superposed the normalized Kretschmann invariant $\bar{\mathcal{K}}$, eq. \eqref{eq:kretschmann}, on the horizon. As \cite{Lehner:2010pn,Lehner:2011wc} already described, the local geometry of the horizon can be interpreted as a dynamical sequence of spherical black holes connected by black strings and our simulations confirm their results. The black string segments are locally GL unstable, leading to a self-similar cascade of instabilities happening on different scales. Indeed, the fact that $\bar{\mathcal{K}}$ is essentially $1$ on the string-like portions of the AH whilst it approaches $6$ near the equator of the bulges corresponding to different generations confirms the picture for the various $L$'s.  On the other hand, the regions near the bulges, where $\bar{\mathcal{K}}$ differs significantly from these two values, correspond to the highly dynamical regions of the AH. In these neck regions, the horizon is becoming thinner on a fast time scale due to mass being accreted by the neighboring bulge and the surrounding string segment. Note that the fact that normalized Kretschmann approaches $1$ on the string segments implies that the spacetime curvature invariant is blowing up like $~1/R^4$, where $R$ the thickness of the horizon.   It is apparent from Fig. \ref{fig:AH_embedding} that in our simulations, the distribution of the higher generation bulges is completely symmetric with respect to the second generation bulge for all $L$'s. This is the expectation for our choice of initial data and given that the modes that govern the dynamics beyond the first generation have this symmetry.

%%%%%%%%%%%%%% L=10, table 1 %%%%%%%%%%%%%%
\begin{table}[h]
    \centering
    \begin{tabular}{ccccccc}
    \hline
    \hline
         & Gen. &$t_i/r_0$  &$n_s$    &$R_{s,i}/r_0$  &$R_{h,f}/r_0$  &$L_{s,i}/R_{s,i}$ \\
         \hline
         & 1        & 69      & 1       &  1          & 2.04         & 10 \\
         \hline
         & 2        & 117     & 1       &  0.0586     & 0.299        & 136 \\
         \hline
         & 3        & 122.8   & 1       &  0.0343     & 0.124        & 109 \\
         \hline
         & 4        & 126.47 & $\geq$1 &  0.01       & ?            & 264\\
    \hline
    \hline
    \end{tabular}
    \caption{Properties of the generations for the $L=10$ case using the definitions in \cite{Lehner:2010pn,Lehner:2011wc}. By comparing the results for different resolutions, we estimate the errors to be at the 1\% level or smaller for the first and second generations. }
    \label{tab:L10defsLP}
\end{table}

In the following we consider in detail the dynamics of the formation of the various generations. The definition of when the various generations have formed is somewhat arbitrary because the unstable black string is continuing to evolve and no portion of the AH is genuinely stationary.  In some cases, a given generation of bulges has not had time to fully form by the time our simulation ends; in other cases, the dynamics of the generations becomes so complex that some quantities are ambiguous. See the  detailed discussion below. In either situation, some geometric quantities of the generations cannot be meaningfully measured according to the definitions below. When this happens, we indicate it with the  symbol  ``?'' in Tables \ref{tab:L10defsLP} and \ref{tab:gens_all}.  

Refs. \cite{Lehner:2010pn,Lehner:2011wc} defined the formation of a new generation as the time when a newly forming bulge has a radius which is 1.5 times the radius of the surrounding string segment. For each generation, one measures the number $n_s$ of satellite black holes that form per string segment, their radii $R_{h,i}$ as well as the radii $R_{s,i}$ of the string segments and their respective lengths $L_{s,i}$. Using these definitions, we summarize our results for the $L=10$ case in Table \ref{tab:L10defsLP}. This table can be directly compared with Table I in \cite{Lehner:2010pn}. 

We find reasonably good agreement between our results and \cite{Lehner:2010pn}, within the errors, up to the second generation; we note that apparent disagreement in the thickness of the string segment $R_{s,2}/r_0$ may be attributed to the rather ambiguous definition of $R_{s,2}$ since the string is slightly non-uniform beyond the first generation; this lack of non-uniformity of the string-like portions of the horizon becomes more exacerbated as the evolution proceeds, see the bottom two panels in Fig. \ref{fig:2nd_gen}. On the other hand, the agreement on the equatorial radius of the first and second generation bulges $R_{h,f}$ is very good. Beyond the second generation, the agreement between our results and those in \cite{Lehner:2010pn} does not seem to be that good anymore; \cite{Lehner:2010pn} mention large errors in their simulatons at this point, which could explain the apparent discrepancies. 

With these definitions of the generations, \cite{Lehner:2010pn} suggested, based on their data, that the instabilities that give rise to the third and higher generations unfold on a time scale which is approximately $1/4$ of the preceding one. We do not find evidence in our data of a global time scale relating the subsequent generations in Table \ref{tab:L10defsLP} for $L=10$ or in Table \ref{tab:gens_all} for any of the values of $L$ that we have considered. In fact, we will argue below there cannot be such a global time scale and that the development of the instabilities beyond the second generation is a rather chaotic process that depends on the local dynamics of the bulges and the string segments, which in turn is dictated by the gravitational self-attraction and local tension of the string.

\begin{figure}
    \centering
    \includegraphics[width=0.48\textwidth]{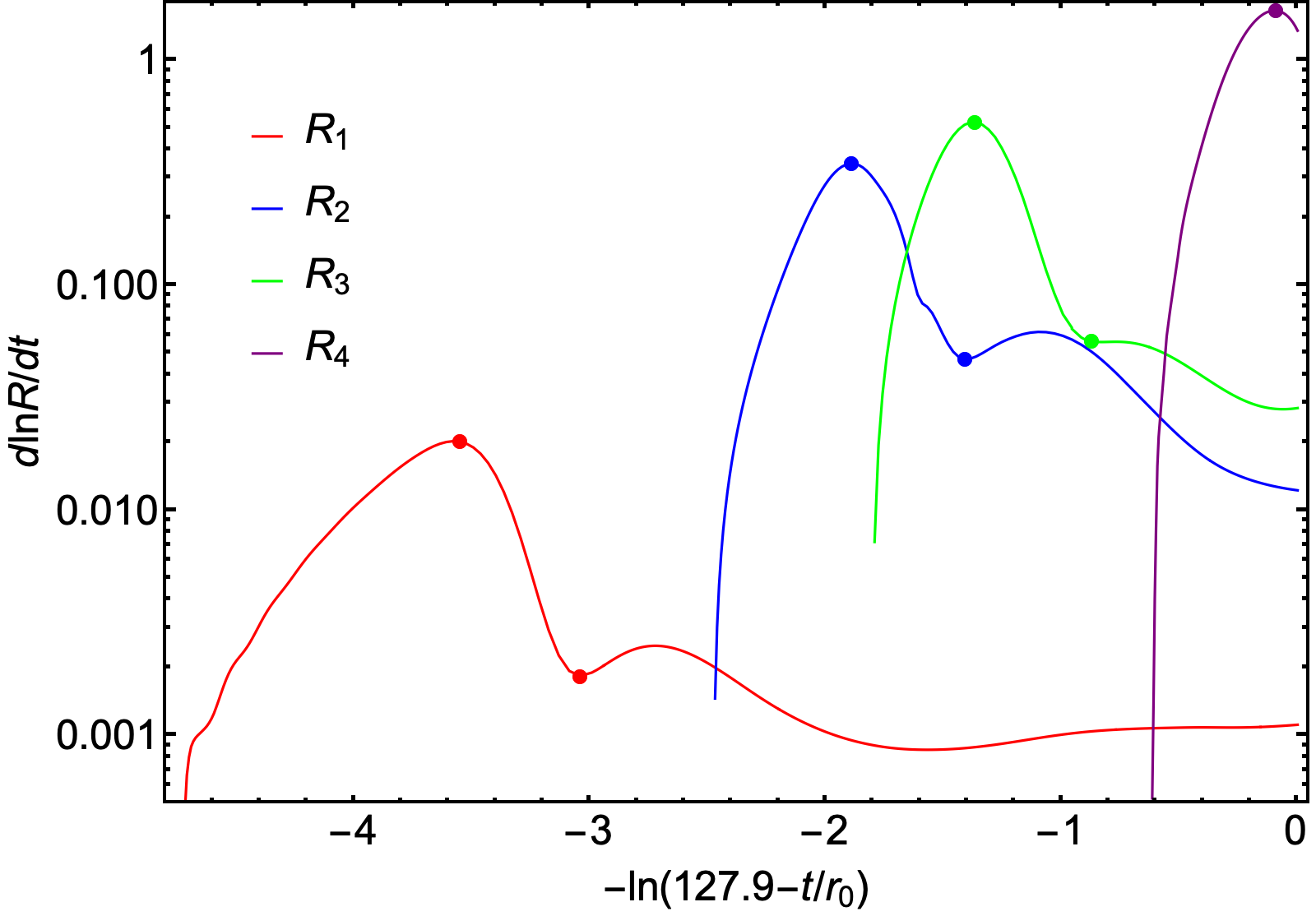}
    \caption{Time derivative of the logarithm of the equatorial radius for each of the generations for the $L=10$ case as functions of time.}
    \label{fig:dlnRdt}
\end{figure}

To discuss our results, we found it useful to use slightly different definition for the time of formation of the generations. The motivation is the following. In Fig. \ref{fig:dlnRdt} we plot the time derivative of the logarithm of the radius of the bulges corresponding to the first four generations as functions of time for the $L=10$ black string; other values of $L$ give similar plots. The absolute maxima in this plot, marked with a dot, indicate the time $t_{p,i}$ at which the growth of the $i$-th generation is fastest compared to its size; after this point, the rate of change of the radius of the $i$-th generation bulge slows down until it develops a local minimum, also marked with a dot in Fig. \ref{fig:dlnRdt}. We choose this local minimum as the time $t_{n,i}$ of formation of the $i$-th generation and we measure its geometric properties at this point.  After this time, the radius of the bulge exhibits some non-trivial (yet slow) dynamics due to the interactions with nearby bulges from the preceding or subsequent generations. The definitions of $R_{h,f}$ and $R_{s,i}$ remain unchanged. We collect the properties of the different generations that we have managed to observe according to these definitions for the various values of $L$ in Table \ref{tab:gens_all}. 

%%%%%%%%%%%%%% Gather tables
\begin{table*}[t!]
    \centering
    \begin{tabular}{c|ccccccc}
    \hline
    \hline
       $L/r_0$  & Gen. &$t_{p,i}/r_0$ &$t_{n,i}/r_0$  &$n_s$    &$R_{s,i}/r_0$  &$R_{h,f}/r_0$  &$R_{s,i}/L_{s,i}$ \\
         \hline
         \multirow{3}{*}{$8$}
         & 1   & 163.1         & 174.7         & 1      & 1       & 1.846   & 0.125         \\
         & 2   & 185.3         & 186.6         & 1      & 0.106   &  0.145  & 0.019        \\
         & 3   & 188.1/188.1   & 188.63/188.62 & 2      & 0.0230  &  0.0657/0.0585  & 0.0078 \\
         \hline
         \multirow{4}{*}{$10$}
         & 1   & 93         & 107         & 1       &  1          & 2.04         & 0.1 \\
         & 2   & 121.3      & 123.8       & 1       &  0.162      & 0.299        & 0.023 \\
         & 3   & 124.0      & 125.6       & 1       &  0.0539     & 0.124        & 0.016 \\
         & 4   & 126.8  & ?   & $\geq 1$/1 &  0.0185/0.0165    & ?      & 0.0076/0.012\\
         \hline
         \multirow{3}{*}{$12$}
         & 1   & 78        & 96.4     & 1       & 1      & 2.221     & 0.0833  \\
         & 2   & 111.7     & 115.35   & 1       & 0.209  & 0.457     & 0.025  \\
         & 3   & chaotic   & ?        & $\geq 1$& 0.0452 & ?         & 0.010  \\
         \hline
         \multirow{3}{*}{$16$}
         & 1   & 91           & 112.7    & 1        & 1     & 2.51  & 0.0625    \\
         & 2   &128.6         & 133.7    & 1        & 0.351 & 0.723 & 0.034  \\
         & 3   & chaotic      & ?        & $\geq 1$ & 0.0263  & ?  & 0.0039        \\
         \hline
         \hline
    \end{tabular}
    \caption{Properties of the various generations for different $L$'s according to our definitions for the time of formation.  We estimate the errors to be at 1-5\% level for the 3rd generation and beyond, while they are $\lesssim 1\%$ for the first and second generations. There is no global time scale relating the time of of formation of the generations beyond the second one. }
    \label{tab:gens_all}
\end{table*}

\begin{figure}[t]
    \centering
    \includegraphics[width=0.45\textwidth]{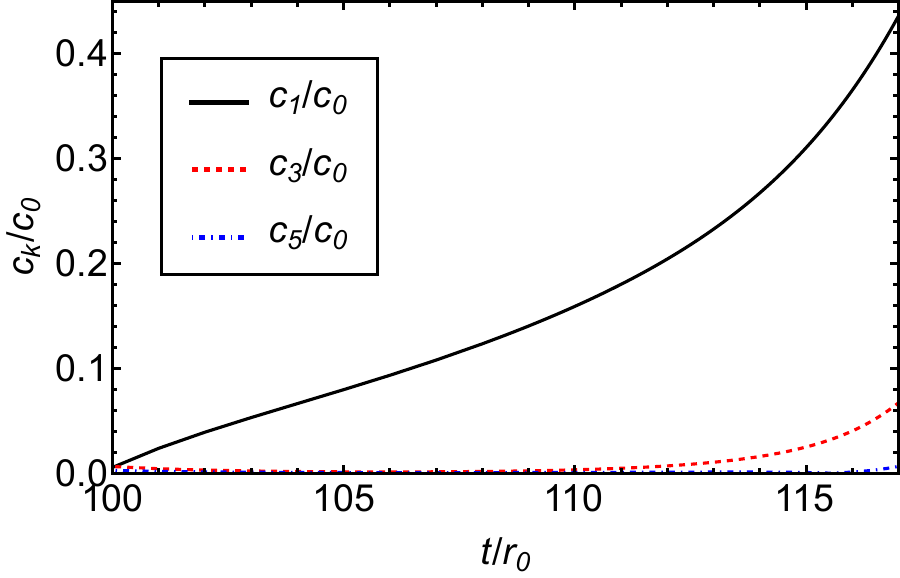}
    \caption{First few Fourier coefficients of the AH radius $R(t,z)$ as functions of time on the first generation string segment. The first Fourier coefficient $c_1$ dominates the dynamics. The even Fourier coefficients are negligible compared to the odd ones.}
    \label{fig:modes_2nd_gen}
\end{figure}

We find that up until the formation of the third generation, the dynamics is governed by the symmetry of the problem. The location along the string where the first bulge appears depends on the initial data; once formed, the first generation bulge is surrounded by a uniform string that eventually becomes unstable, leading to the formation of a second generation bulge. Given that there is no net momentum along the string because of our choice of initial data, if the first GL harmonic is the one driving the formation of the second generation bulge (as in all cases that we have considered, see Fig. \ref{fig:modes_2nd_gen}), the latter has to form at the centre of the preceding segment of string by symmetry. Fig. \ref{fig:2nd_gen} shows that this is precisely what happens. Following \cite{Lehner:2011wc}, we decompose the string radius $R(t,z)$ in the coordinate region $z\in[0,5r_0]$ covering the first string segment for the $L=10$ case,
\begin{equation}
    R(t,z)=c_0+\sum_{k=1}^{\infty} c_k\textstyle\sin\left(\frac{\pi k z}{L_{s,1}}\right)\,,
\end{equation}
to extract the Fourier coefficients $c_k$ that capture the ``strength" of the various GL harmonics. Fig. \ref{fig:modes_2nd_gen} shows the values of $c_k/c_0$ as functions of time for the first few $k$'s. Just as  \cite{Lehner:2011wc} already found, the odd coefficients dominate and $c_1$ is the one governing the evolution, which is consistent with the formation of a single second generation bulge at the center of the string segment.  For the larger values of $L$ we find that the higher (odd) Fourier coefficients, namely $c_3$ and $c_5$, are ``stronger" compared to $c_1$. This is consistent with the fact that the unstable string segment is longer, so higher harmonics can also be excited. 

\begin{figure}[t]
    \centering
    \includegraphics[width=0.48\textwidth]{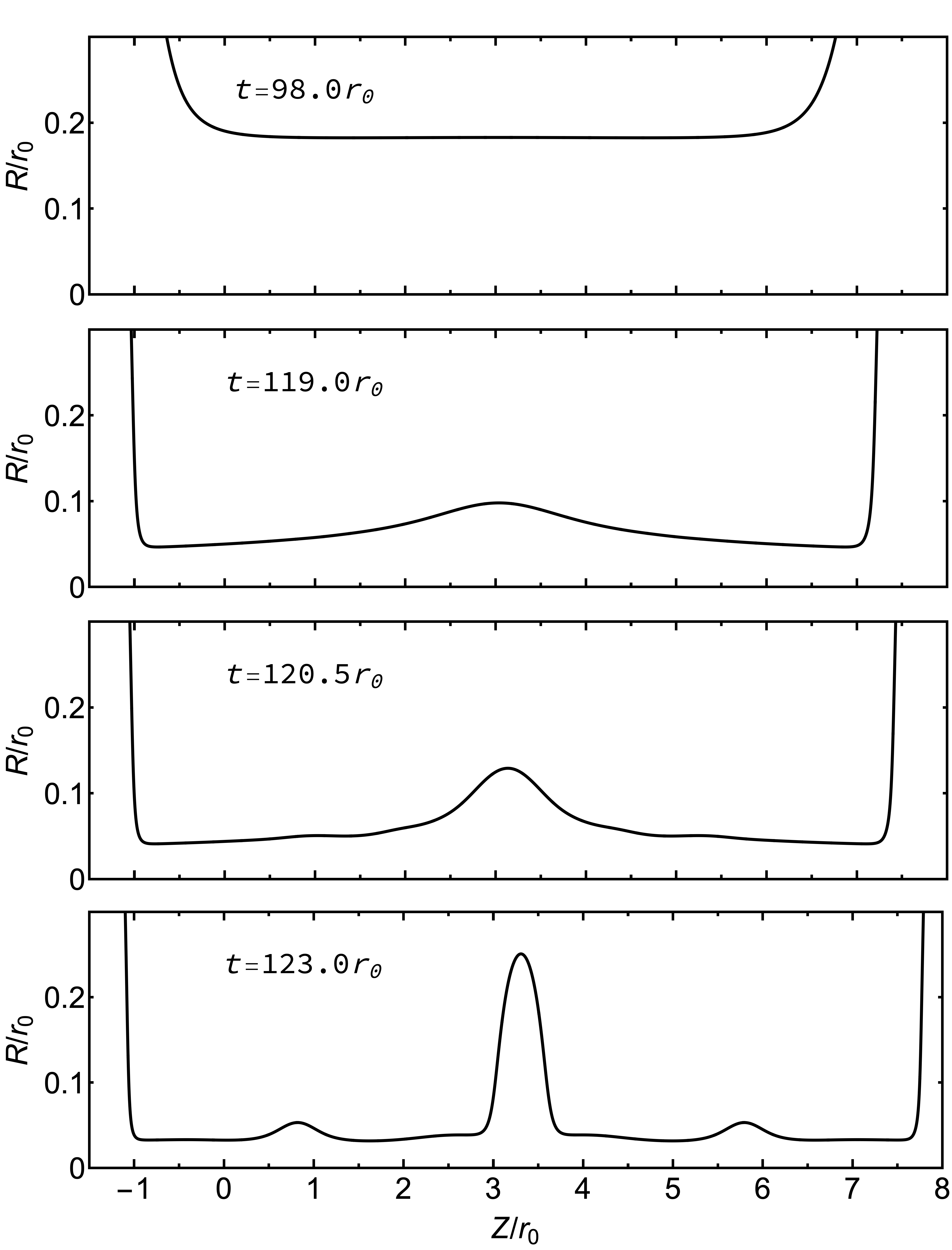}
    \caption{Formation of the second and third generations for the $L=10$ case. We have added a copy of the string to the left of the plots to facilitate the visualization  given that the $Z$ direction is periodic. The second generation forms right at the centre of the preceding string segment; the disturbance that eventually becomes the third generation moves along the string segment at the same time as it grows due to the interactions with the first and second generation blobs. }
    \label{fig:2nd_gen}
\end{figure}

Symmetry also dictates that second generation bulge that forms has zero net velocity along the string direction. From Fig. \ref{fig:2nd_gen}, we see that the second generation bulge is surrounded by two identical segments of (slightly non-uniform) string joining it with the first generation bulge and hence it is in local equilibrium. The disturbances along the string created by the formation of the second generation seed the instabilities that eventually lead to the formation of the third generation, see third panel in Fig. \ref{fig:2nd_gen}. It seems that this is the mechanism through which subsequent generations form from the preceding ones.\footnote{Presumably this is also how the mode that eventually leads to the formation of the second generation gets excited, but it is harder to see from our data.} The string segments surrounding the second generation eventually become unstable but the formation of the third generation bulges is affected by the presence of two unequal bulges at the respective ends in the following way. As the third panel in Fig. \ref{fig:2nd_gen} shows, the disturbance that eventually becomes a third generation bulge is surrounded by string segments of unequal thickness because the bulges at the respective ends have different sizes. Because of this non-uniformity of the string segment and the unequal sizes of the bulges, the disturbance no longer appears at the centre of the preceding string; consequently, as it grows it is also attracted towards the first or second generation bulge depending on its initial location along the string.  Therefore, as the third generation forms by accreting mass from the surrounding string segments, the growing bulge acquires a certain net velocity towards one of the preceding generation bulges, leading to a further thinning of one of the surrounding string segments. In the example shown in Fig. \ref{fig:2nd_gen}, the third generation bulge moves towards the first generation one. This local motion of the bulges that inevitably arises at the third generation and beyond interferes with the development of the local GL instabilities because it leads to a further thinning of certain string segments and it occurs on the same time scale as the formation of the generations themselves. Clearly this pattern of local bulges' motion and instabilities taking place at the same time is what determines the evolution of the system beyond the second generation. Ultimately, this is the reason why from the third generation onwards there cannot be a global time scale relating the subsequent generations that form as a result of local GL instabilities. Instead,  the process becomes rather chaotic. It would be interesting to quantify the latter more precisely.

\begin{figure}[t]
    \centering
    \includegraphics[width=0.48\textwidth]{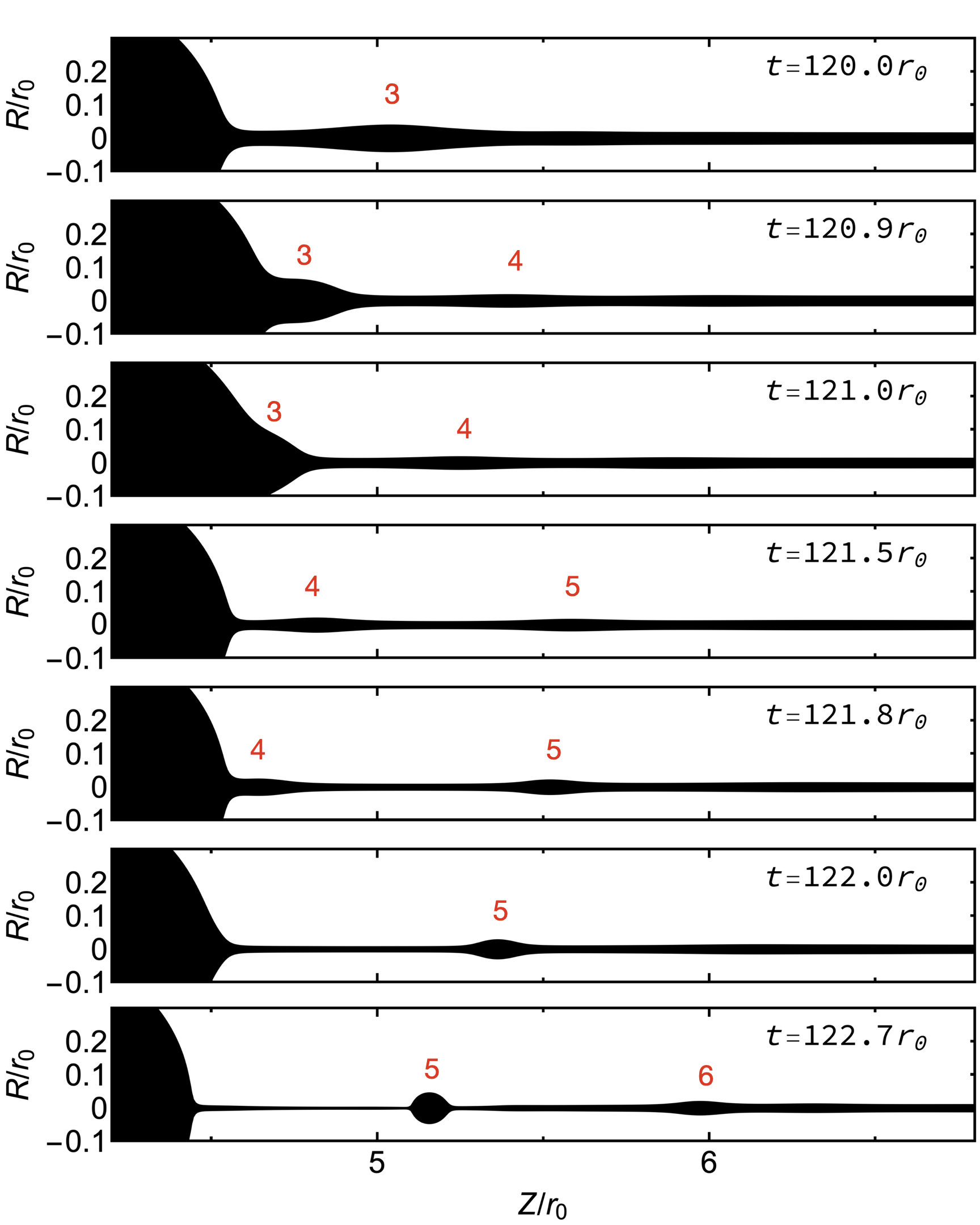}
    \caption{Local dynamics of the bulges for the $L=12$ case. The numbers in red label the generations. The third and fourth generation bulges do not have enough time to fully form before being absorbed by the existing second generation bulge (big blob on the left). On the other hand,  the fifth generation bulge fully forms, even though it migrates towards the second generation bulge as it forms.}
    \label{fig:bulge_dynamics_L12}
\end{figure}

We illustrate in detail the local dynamics of the bulges for the $L=12$ case in Fig. \ref{fig:bulge_dynamics_L12}. The third generation bulge starts forming near the second generation bulge (first panel) and it moves towards the latter as it grows. However, it does not have time to fully form before being absorbed (second and third panels). A would-be fourth generation bulge then starts forming, but again it is absorbed by the second generation bulge before it has had time to fully form; at the same time, the fifth generation bulge starts to form, but in this case it is located far enough from the second generation bulge so that it has enough time to fully form. One can see from the lower two panels in Fig. \ref{fig:bulge_dynamics_L12} that the fifth generation bulge moves towards the second generation one while the sixth generation is just beginning to form. In the $L=12$ case we find that the minimum thickness of the string is located near where the string segment joins the fifth generation bulge.

\begin{figure}[t]
    \centering
    \includegraphics[width=0.48\textwidth]{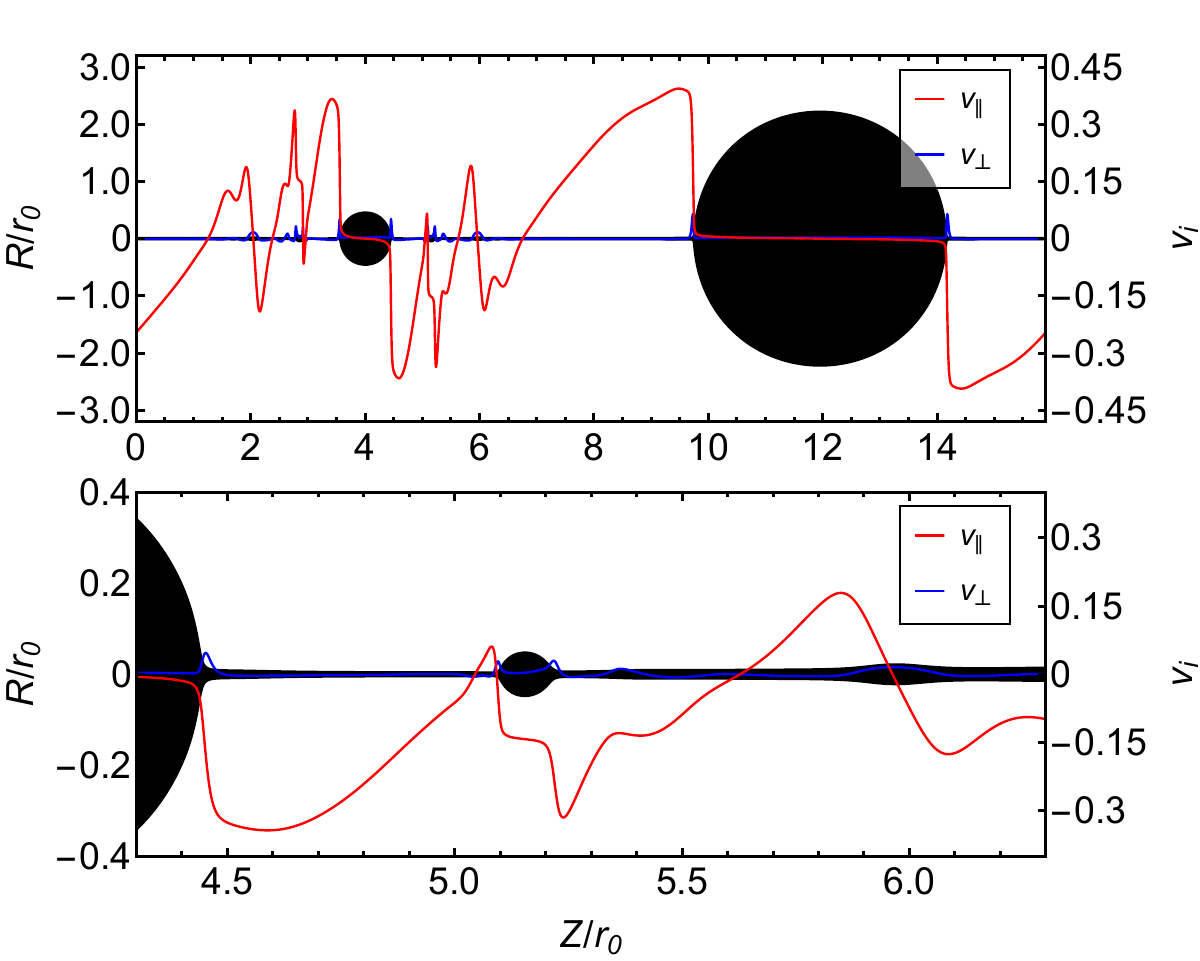}
    \caption{\textit{Top}: Tangential and orthogonal velocities of a null ray co-moving with the AH in the final stages of the evolution of the unstable black string with $L=12$. \textit{Bottom}: Zoom in of the thinnest part of the string, showing the second generation blob (left) and the fifth generation blob (center). The latter is moving towards the former.}
    \label{fig:loc_velocities}
\end{figure}

In order to get a qualitative picture of the motion of the bulges at the last stages of our simulations,  we consider a particular null ray co-moving with the AH,
\begin{equation}
V=\partial_t+\dot z\,\partial_z + \left(\dot x+h'\,\dot z\right)\partial_x\,,
\label{eq:null_vec}
\end{equation}
where the dot $\dot\,$ denotes the derivative with respect to the parameter $t$, and $\dot z$ and $\dot x$  correspond to the tangential ($v_\parallel$) and orthogonal ($v_\perp$) velocities of the null ray with respect to asymptotic observers at rest; these velocities provide some measure  of how the various portions of the AH are moving with respect to a frame that is at rest with respect to asymptotic infinity.\footnote{In practice, $\dot z$ and $\dot x$ are found by requiring that \eqref{eq:null_vec} is null and that the quadratic equation $g_{\mu\nu}V^\mu V^\nu=0$ has zero discriminant.} In Fig. \ref{fig:loc_velocities} we plot the tangential and orthogonal velocities of  such a light ray that is co-moving with the AH at the last stage of our simulation. This figure   is consistent with the local dynamics of the blobs described in the previous paragraphs since it shows that while the smaller blobs accrete matter from the surrounding segments of string, as shown by the local non-zero orthogonal velocities, they are collectively being dragged towards the second or first generation blobs. Furthermore, the tangential velocities are relativistic and they are much larger than the orthogonal ones. However, the motion of the blobs is slow compared to the development of the local GL instabilities on smaller scales.

The number of bulges that form per generation and their location along the string depends on the value of $L$ but the main features of their local dynamics are essentially the same for all cases. The motion of the bulges interferes with their growth and hence their time of formation as well as with the thicknesses of the resulting segments of black string surrounding them. Because of this local motion, some black string segments become thinner than they would otherwise be solely due to the mass accretion of a nearby bulge that is forming. Since it is the thickness of the local string segment what determines the growth rates of the local GL instabilities and the harmonics that will drive the formation of the the next generations, the local motion accelerates the approach to the pinch off. Therefore,  while there is no global time scale relating subsequent generations beyond the second one, the fate of the black string is sealed once the first GL instability sets in: the black string will necessarily pinch off in finite asymptotic time. Furthermore, the process accelerates as it approaches the singularity. We should emphasize that the special role that the third generation seems to play in this discussion is an artefact of our choice of initial data; other initial conditions, for instance with a net momentum along the string, could induce the local bulge dynamics  at a different stage in the evolution and the third generation would no longer be singled out.

%%%%%%%%%%%%%%%%%%%%%%%
\subsection{Approach to the singularity}
\label{sec:singularity}
%%%%%%%%%%%%%%%%%%%%%%%

The main result of \cite{Lehner:2010pn,Lehner:2011wc} was to convincingly demonstrate that the black string pinches off in finite asymptotic time. In this section we corroborate this result for the various $L$'s that we have considered. 

As previously observed \cite{Lehner:2010pn,Lehner:2011wc,Figueras:2017zwa}, the minimum of the areal radius of the black string follows an approximate scaling law,
\begin{equation}
    R_\text{min} = a\,(t-t_c)\,,
    \label{eq:scaling}
\end{equation}
where $a$ is a dimensionless proportionality constant and $t_c$ is the pinch off time. By fitting our numerical data to \eqref{eq:scaling} we can extract the values of $a$ and $t_c$ for the different $L$'s. For the $L=8,16$ cases we did not manage to get as close to the pinch off as in the $L=10,12$ cases due to the lack of sufficient computational resources.  

\begin{figure}[t]
    \centering
    \includegraphics[width=0.48\textwidth]{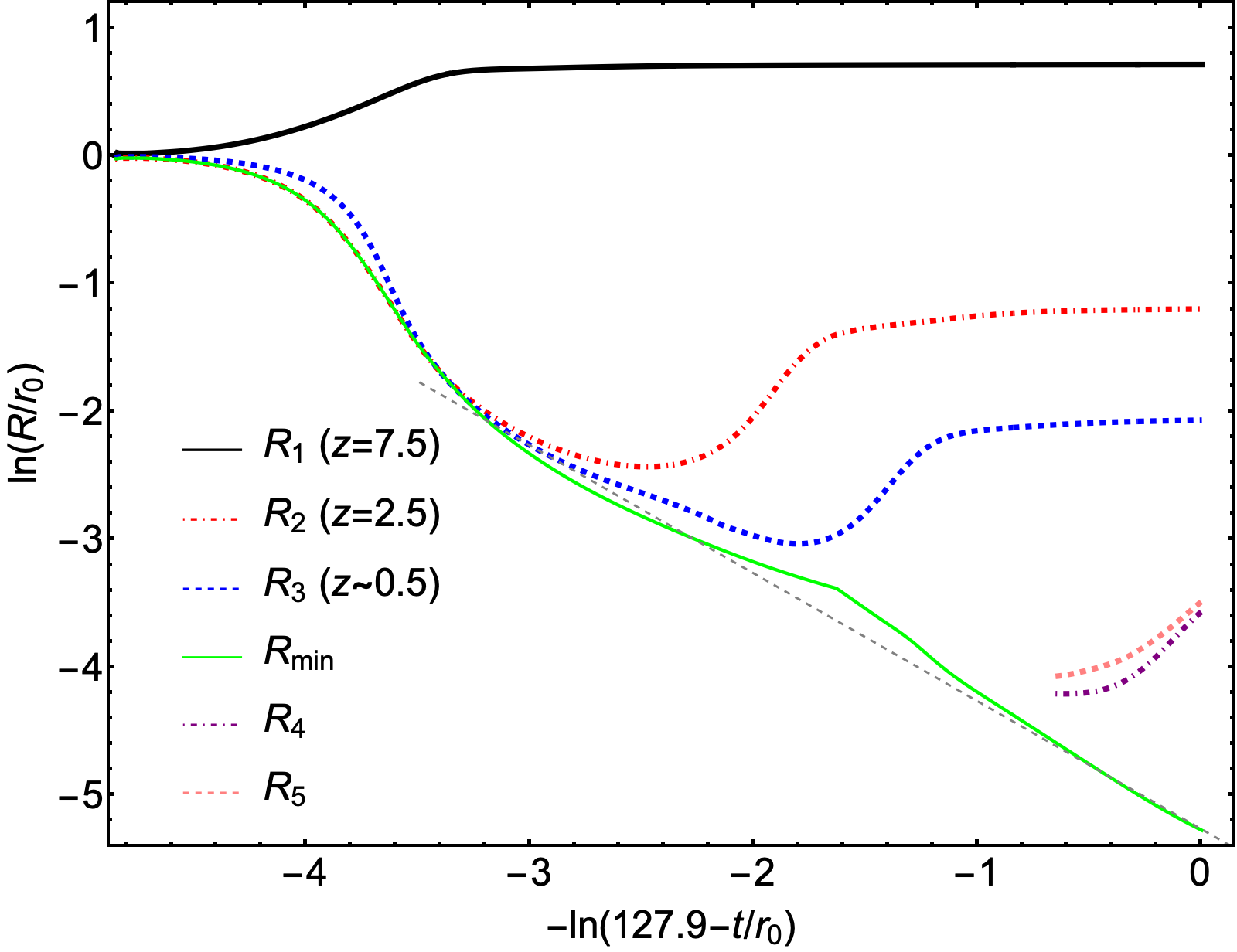}
    \caption{Logarithm of the areal radius $R$ of the AH vs the logarithm of the time to the singularity for the $L=10$ case. We display the evolution of the minimum string radius together with the equatorial radii of several generations. Other $L$'s give qualitatively similar plots.}
    \label{fig:lnR_vs_lnt}
\end{figure}

In Fig. \ref{fig:lnR_vs_lnt} we display the logarithm of the minimum areal radius of the string as a function of the logarithm of the time to the singularity for the $L=10$ case; this figure can be compared with Fig. 4 in \cite{Lehner:2010pn}. We collect the results of the fits for the various $L$'s in Table \ref{tab:scaling}.

\begin{table}[h]
    \centering
    \centering
    \begin{tabular}{c|c|c}
    \hline
    \hline
          $L/r_0$ & $a$  & $t_c/r_0$    \\
         \hline
         8        & 0.0058     & 189.6      \\
         \hline
          10        & 0.0052  & 127.9      \\
         \hline
          12       & 0.0048  & 123.5         \\
         \hline
          16       & 0.0045 & 148.2  \\
    \hline
    \hline
    \end{tabular}
    \caption{Slope of the scaling law \eqref{eq:scaling} and pinch off time for the different $L$'s.}
    \label{tab:scaling}
\end{table}

The results in Table \ref{tab:scaling} suggest that the slope of the scaling law \eqref{eq:scaling} is independent of $L$, at least within the errors and taking into account that we did not manage to get equally close to the singularity for all $L$'s. If confirmed, this could be interpreted as providing evidence for the existence of a universal scaling solution governing the pinch off of the black string. It is also interesting to note that the values of the slope that we find for the black string are essentially half of the values that the authors of \cite{Figueras:2017zwa} found in the case of the ultraspinning instability of Myers-Perry black holes in six spacetime dimensions. In the latter case, the local geometry of the AH is that of a black membrane. It would be interesting to understand the origin of this factor of two difference between the respective scaling laws.

%%%%%%%%%%%%%%%%%%%%%%%
\section{Discussion}
\label{sec:discussion}
%%%%%%%%%%%%%%%%%%%%%%%

In this article we have reproduced and confirmed the famous results of Lehner and Pretorius \cite{Lehner:2010pn} on the endpoint of the GL instability of black strings in five spacetime dimensions. In particular, we provided further evidence that the unstable black string evolves into a self-similar sequence of uniform strings connecting spherical black holes on different scales that should result in a pinch off in finite asymptotic time. Furthermore, we have extended the results of \cite{Lehner:2010pn} in several directions: First, we have considered unstable black strings of different lengths and equal thicknesses, and shown that the approach to the singularity is qualitatively the same as in the case considered in \cite{Lehner:2010pn}. Second, our numerical simulations have allowed us to confidently get closer to the singularity than before, reaching up to four or five generations in the $L=10$ and $L=12$ cases respectively, whilst preserving the $\mathbb Z_2$ symmetry about the second generation blob. This symmetry  and the particular role that the third generation plays in the evolution are a consequence of the type of initial conditions that we (and \cite{Lehner:2010pn}) have considered.

While our results and those of \cite{Lehner:2010pn} agree very well up to and including the second generation, we have seen no evidence of a global timescale governing the formation of the higher generations. We have argued that such a timescale cannot exist because of the local motion of the higher generation bulges caused by the tension and gravitational self-attraction of the string. Such a motion takes place on a timescale which is of the same order as the time of formation of the next generation, and it leads to further thinning of certain string segments, thus speeding up the approach to the singularity. However, the basic picture remains; namely, the minimum thickness of the string follows a scaling law \eqref{eq:scaling}, and the timescales for the development of local GL instabilities, which are determined by the thickness of the string segments, are shorter than the time of formation of the singularity, at least for the duration of our simulations.

There are several directions for future work. One of the most interesting open questions about the evolution of the GL instability of the black string is the nature of the singularity at the pinch off. This has far reaching implications for the physical consequences of certain violations of the WCCC and the predictivity of general relativity as a classical theory of gravity. Arguably, the black string is the cleanest system where these questions may be addressed. In non-relativistic incompressible fluids, the work of \cite{Eggers:1993aa} showed that the pinch off of an axisymmetric column of fluid is governed by a universal scale invariant attractor solution of the Navier-Stokes equations. The solution of \cite{Eggers:1993aa} shows that essentially only microscopic a region of the fluid is involved in the pinch off and, furthermore, this region is insensitive to the evolution of the fluid on macroscopic scales. This implies that the loss of predictivity of the classical hydrodynamic equations at the pinch off is minimal and the details of how the pinch off takes place are irrelevant for the macroscopic evolution of the fluid. Ref. \cite{Emparan:2020vyf} conjectured that a similar picture would hold in the case of the GL instability of black strings. A related question is whether an infinite number of generations or only a finite number of them form before the singularity; in the former case, the horizon would develop a fractal structure, while in the latter case this fractal structure would break down at some point.\footnote{We thank Frans Pretorius for pointing this out to us.} We hope to report on the nature of the pinch off in future work.   

Other important and related open questions are whether null infinity is complete and determine the location of the event horizon (EH). Given that the most of the geometry of the AH of the unstable black string can be accurately described as a sequence of quasistationary black strings connecting black holes, one would expect that the AH is very close to the EH almost everywhere. Ref. \cite{Choptuik:2003qd} confirmed this at least for the initial stages of the evolution. However, the pinch off is expected to take place in a highly dynamical region of the geometry of the black string, precisely where the AH and the EH may differ substantially. One can locate the EH by shooting null geodesics backwards in time \cite{Anninos:1994ay,Libson:1994dk}, starting from the last snapshot of the simulation. However, in our case, this approach poses a serious practical difficulty because the size of the checkpoint files that we have to store increases exponentially close to the pinch off and we quickly run out of storage space in a normal cluster.

As the evolution of the black string approaches the pinch off, one would expect that eventually higher derivative corrections to the Einstein-Hilbert Lagrangian would become important. It would be interesting to quantify whether this is indeed the case in the regime that can be probed with numerical simulations and how such corrections affect the dynamics. In a recent breakthrough, Refs. \cite{Kovacs:2020pns,Kovacs:2020ywu} (see also \cite{AresteSalo:2022hua}) have shown that certain higher derivative theories of gravity in higher dimensions are well-posed in reasonably straightforward modifications of the gauges commonly used in numerical relativity. Therefore, it should be possible to study higher derivative corrections to the evolution of the GL instability. Work to address this question is in progress.

\section*{Acknowledgements}

We would like to thank Roberto Emparan, Luis Lehner and Frans Pretorius for very useful comments and discussions on a draft of this article. We also want to thank the entire GRChombo\footnote{\texttt{www.grchombo.org}} collaboration, and Miren Radia in particular, for their support and code development work. PF would like to thank the Enrico Fermi Institute and the Department of Physics of the University of Chicago for hospitality during the final stages of this work. This work was first presented by one us at the GR23 conference, $3^\text{th}$--$8^\text{th}$ of July 2022 and at the workshop ``Fundamental Aspects of Gravity", Imperial College London, $8^\text{th}$--$12^\text{th}$ of August 2022. We would like to thank the organizers for the invitation, the participants for stimulating discussions and the ``Gravity Theory Trust" for its support. We CG and PF were supported by the European Research Council Grant No. ERC-2014-StG 639022-NewNGR. PF is supported by Royal Society University Research Fellowship Grants No. UF140319, RGF\textbackslash EA\textbackslash 180260, URF\textbackslash R\textbackslash 201026 and RF\textbackslash ERE\textbackslash 210291. TF is supported by a PhD studentship from the Royal Society RS\textbackslash PhD\textbackslash 181177.  The simulations presented used PRACE resources under Grant No. 2020235545, PRACE DECI-17 resources under Grant No. 17DECI0017, the CSD3 cluster in Cambridge under projects DP128 and DP214. The Cambridge Service for Data Driven Discovery (CSD3), partially operated by the University of Cambridge Research Computing on behalf of the STFC DiRAC HPC Facility. The DiRAC component of CSD3 is funded by BEIS capital via STFC capital Grants No. ST/P002307/1 and No. ST/ R002452/1 and STFC operations Grant No. ST/R00689X/1. DiRAC is part of the National e-Infrastructure.\footnote{\texttt{www.dirac.ac.uk}} The authors gratefully acknowledge the Gauss Centre for Supercomputing e.V.\footnote{\texttt{www.gauss-centre.eu}} for providing computing time on the GCS Supercomputer SuperMUC-NG at Leibniz Supercomputing Centre.\footnote{\texttt{www.lrz.de}} Calculations were performed using the Sulis Tier 2 HPC platform hosted by the Scientific Computing Research Technology Platform at the University of Warwick. Sulis is funded by EPSRC Grant EP/T022108/1 and the HPC Midlands+ consortium. This research also utilised Queen Mary’s Apocrita HPC facility, supported by QMUL Research-IT \cite{apocrita}. For some computations we have also used the Young Tier 2 HPC cluster at UCL; we are grateful to the UK Materials and Molecular Modelling Hub for computational resources, which is partially funded by EPSRC (EP/P020194/1 and EP/T022213/1). The authors also thankfully acknowledge the computer resources at MareNostrum and the technical support provided by BSC (RES-FI-2020-3-0007). This research also utilised Athena at HPC Midlands+ cluster. This work also used the ARCHER2 UK National Supercomputing Service under project E775.\footnote{\texttt{www.archer2.ac.uk}}

\appendix

\section{Convergence tests}
\label{app:convergence}

In this appendix we provide the results of the convergence tests. We monitor the AH area since it plays a fundamental role in the interpretation of the results and, in turn, it is sensitive to the overall structure of the horizon, thus providing a good idea of the accuracy level of the simulations. To carry out the tests we evolved the $L=10$, $r_0=1$ black string across different resolutions. 

\begin{figure}[t!]
    \centering
    \includegraphics[width=0.48\textwidth]{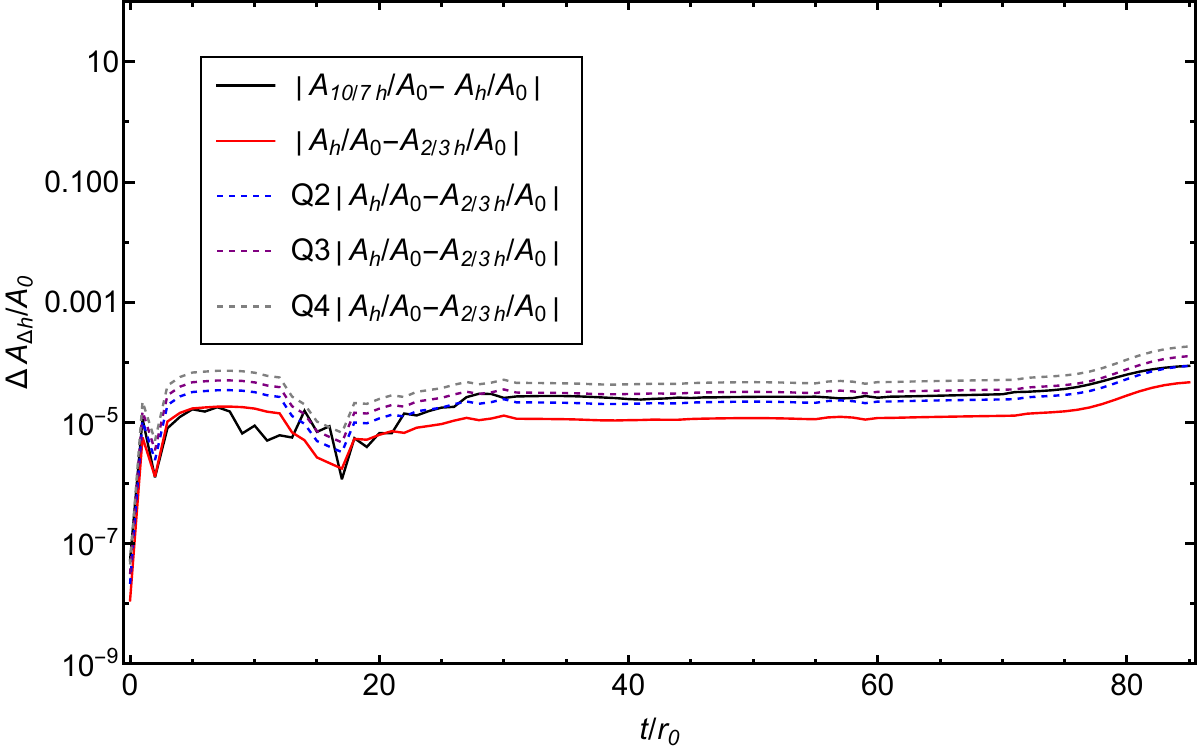}
    \caption{Convergence test for the AH area for the $r_0=1$, $L=10$ black string. The medium resolution run has spacing $h=0.25r_0$ in the coarsest grid level. This plot shows that the order of convergence is roughly three. The computational cost of the high resolution run limited the time of the tests. }
    \label{fig:ah_area_conv}
\end{figure}

The medium resolution run has a coarsest grid spacing of $\Delta_\text{med}=h=0.25r_0$ and this is the typical resolution that we have used to present the results in the paper. The low and high resolution runs have grid spacings $\Delta_\text{low}=\frac{10}{7}h$ and $\Delta_\text{high}=\frac{2}{3}h$ respectively. The reason for these choices is the following. The low resolution cannot be too low, e.g., $\Delta_\text{low}=2h$, because otherwise the code would crash shortly after the formation of the first generation; on the other hand, too high resolution is unfeasible with our limited computational resources. Therefore, the chosen grid spacings for the tests seem to be a good compromise.

In Fig. \ref{fig:ah_area_conv} we display the results of the convergence tests. In this plot, we compare the AH area computed with different resolutions in terms of the convergence factor
\begin{equation}
Q_n=\frac{\Delta_\text{low}^n-\Delta_\text{med}^n}{\Delta_\text{med}^n-\Delta_\text{high}^n}\,.
\end{equation}
Due to the high computational cost of the high resolution run, we could only evolve it for a limited time, which in turn limits the evolution time of the tests. This plot shows that the order of convergence is roughly 3 throughout the evolution. This is expected since even though we use a method that is $6^\text{th}$ order in space and $4^\text{th}$ order in time, the interpolation/extrapolation at the level boundaries typically reduce the order of convergence.

\bibliography{refs}

\begin{thebibliography}{62}%
\makeatletter
\providecommand \@ifxundefined [1]{%
 \@ifx{#1\undefined}
}%
\providecommand \@ifnum [1]{%
 \ifnum #1\expandafter \@firstoftwo
 \else \expandafter \@secondoftwo
 \fi
}%
\providecommand \@ifx [1]{%
 \ifx #1\expandafter \@firstoftwo
 \else \expandafter \@secondoftwo
 \fi
}%
\providecommand \natexlab [1]{#1}%
\providecommand \enquote  [1]{``#1''}%
\providecommand \bibnamefont  [1]{#1}%
\providecommand \bibfnamefont [1]{#1}%
\providecommand \citenamefont [1]{#1}%
\providecommand \href@noop [0]{\@secondoftwo}%
\providecommand \href [0]{\begingroup \@sanitize@url \@href}%
\providecommand \@href[1]{\@@startlink{#1}\@@href}%
\providecommand \@@href[1]{\endgroup#1\@@endlink}%
\providecommand \@sanitize@url [0]{\catcode `\\12\catcode `\$12\catcode
  `\&12\catcode `\#12\catcode `\^12\catcode `\_12\catcode `\%12\relax}%
\providecommand \@@startlink[1]{}%
\providecommand \@@endlink[0]{}%
\providecommand \url  [0]{\begingroup\@sanitize@url \@url }%
\providecommand \@url [1]{\endgroup\@href {#1}{\urlprefix }}%
\providecommand \urlprefix  [0]{URL }%
\providecommand \Eprint [0]{\href }%
\providecommand \doibase [0]{https://doi.org/}%
\providecommand \selectlanguage [0]{\@gobble}%
\providecommand \bibinfo  [0]{\@secondoftwo}%
\providecommand \bibfield  [0]{\@secondoftwo}%
\providecommand \translation [1]{[#1]}%
\providecommand \BibitemOpen [0]{}%
\providecommand \bibitemStop [0]{}%
\providecommand \bibitemNoStop [0]{.\EOS\space}%
\providecommand \EOS [0]{\spacefactor3000\relax}%
\providecommand \BibitemShut  [1]{\csname bibitem#1\endcsname}%
\let\auto@bib@innerbib\@empty
%</preamble>
\bibitem [{\citenamefont {Abbott}\ \emph
  {et~al.}(2016{\natexlab{a}})\citenamefont {Abbott} \emph
  {et~al.}}]{LIGOScientific:2016aoc}%
  \BibitemOpen
  \bibfield  {author} {\bibinfo {author} {\bibfnamefont {B.~P.}\ \bibnamefont
  {Abbott}} \emph {et~al.} (\bibinfo {collaboration} {LIGO Scientific,
  Virgo}),\ }\bibfield  {title} {\bibinfo {title} {{Observation of
  Gravitational Waves from a Binary Black Hole Merger}},\ }\href
  {https://doi.org/10.1103/PhysRevLett.116.061102} {\bibfield  {journal}
  {\bibinfo  {journal} {Phys. Rev. Lett.}\ }\textbf {\bibinfo {volume} {116}},\
  \bibinfo {pages} {061102} (\bibinfo {year} {2016}{\natexlab{a}})},\ \Eprint
  {https://arxiv.org/abs/1602.03837} {arXiv:1602.03837 [gr-qc]} \BibitemShut
  {NoStop}%
\bibitem [{\citenamefont {Abbott}\ \emph
  {et~al.}(2016{\natexlab{b}})\citenamefont {Abbott} \emph
  {et~al.}}]{LIGOScientific:2016sjg}%
  \BibitemOpen
  \bibfield  {author} {\bibinfo {author} {\bibfnamefont {B.~P.}\ \bibnamefont
  {Abbott}} \emph {et~al.} (\bibinfo {collaboration} {LIGO Scientific,
  Virgo}),\ }\bibfield  {title} {\bibinfo {title} {{GW151226: Observation of
  Gravitational Waves from a 22-Solar-Mass Binary Black Hole Coalescence}},\
  }\href {https://doi.org/10.1103/PhysRevLett.116.241103} {\bibfield  {journal}
  {\bibinfo  {journal} {Phys. Rev. Lett.}\ }\textbf {\bibinfo {volume} {116}},\
  \bibinfo {pages} {241103} (\bibinfo {year} {2016}{\natexlab{b}})},\ \Eprint
  {https://arxiv.org/abs/1606.04855} {arXiv:1606.04855 [gr-qc]} \BibitemShut
  {NoStop}%
\bibitem [{\citenamefont {Penrose}(1965)}]{Penrose:1964wq}%
  \BibitemOpen
  \bibfield  {author} {\bibinfo {author} {\bibfnamefont {R.}~\bibnamefont
  {Penrose}},\ }\bibfield  {title} {\bibinfo {title} {{Gravitational collapse
  and space-time singularities}},\ }\href
  {https://doi.org/10.1103/PhysRevLett.14.57} {\bibfield  {journal} {\bibinfo
  {journal} {Phys. Rev. Lett.}\ }\textbf {\bibinfo {volume} {14}},\ \bibinfo
  {pages} {57} (\bibinfo {year} {1965})}\BibitemShut {NoStop}%
\bibitem [{\citenamefont {Penrose}(1969)}]{Penrose:1969pc}%
  \BibitemOpen
  \bibfield  {author} {\bibinfo {author} {\bibfnamefont {R.}~\bibnamefont
  {Penrose}},\ }\bibfield  {title} {\bibinfo {title} {{Gravitational collapse:
  The role of general relativity}},\ }\href
  {https://doi.org/10.1023/A:1016578408204} {\bibfield  {journal} {\bibinfo
  {journal} {Riv. Nuovo Cim.}\ }\textbf {\bibinfo {volume} {1}},\ \bibinfo
  {pages} {252} (\bibinfo {year} {1969})}\BibitemShut {NoStop}%
\bibitem [{\citenamefont {Geroch}\ and\ \citenamefont
  {Horowitz}(1978)}]{Geroch:1978ub}%
  \BibitemOpen
  \bibfield  {author} {\bibinfo {author} {\bibfnamefont {R.~P.}\ \bibnamefont
  {Geroch}}\ and\ \bibinfo {author} {\bibfnamefont {G.~T.}\ \bibnamefont
  {Horowitz}},\ }\bibfield  {title} {\bibinfo {title} {{Asymptotically simple
  does not imply asymptotically Minkowskian}},\ }\href
  {https://doi.org/10.1103/PhysRevLett.40.203} {\bibfield  {journal} {\bibinfo
  {journal} {Phys. Rev. Lett.}\ }\textbf {\bibinfo {volume} {40}},\ \bibinfo
  {pages} {203} (\bibinfo {year} {1978})}\BibitemShut {NoStop}%
\bibitem [{\citenamefont {Geroch}\ and\ \citenamefont
  {Horowitz}(1979)}]{Geroch:1979uc}%
  \BibitemOpen
  \bibfield  {author} {\bibinfo {author} {\bibfnamefont {R.~P.}\ \bibnamefont
  {Geroch}}\ and\ \bibinfo {author} {\bibfnamefont {G.~T.}\ \bibnamefont
  {Horowitz}},\ }\bibinfo {title} {{Global structure of spacetimes}},\ in\
  \href@noop {} {\emph {\bibinfo {booktitle} {{General Relativity}: {An
  Einstein Centenary Survey}}}}\ (\bibinfo {year} {1979})\ pp.\ \bibinfo
  {pages} {212--293}\BibitemShut {NoStop}%
\bibitem [{\citenamefont
  {Christodoulou}(1999{\natexlab{a}})}]{Christodoulou_1999}%
  \BibitemOpen
  \bibfield  {author} {\bibinfo {author} {\bibfnamefont {D.}~\bibnamefont
  {Christodoulou}},\ }\bibfield  {title} {\bibinfo {title} {On the global
  initial value problem and the issue of singularities},\ }\href
  {https://doi.org/10.1088/0264-9381/16/12a/302} {\bibfield  {journal}
  {\bibinfo  {journal} {Classical and Quantum Gravity}\ }\textbf {\bibinfo
  {volume} {16}},\ \bibinfo {pages} {A23} (\bibinfo {year}
  {1999}{\natexlab{a}})}\BibitemShut {NoStop}%
\bibitem [{\citenamefont {Choptuik}(1993)}]{Choptuik:1992jv}%
  \BibitemOpen
  \bibfield  {author} {\bibinfo {author} {\bibfnamefont {M.~W.}\ \bibnamefont
  {Choptuik}},\ }\bibfield  {title} {\bibinfo {title} {{Universality and
  scaling in gravitational collapse of a massless scalar field}},\ }\href
  {https://doi.org/10.1103/PhysRevLett.70.9} {\bibfield  {journal} {\bibinfo
  {journal} {Phys. Rev. Lett.}\ }\textbf {\bibinfo {volume} {70}},\ \bibinfo
  {pages} {9} (\bibinfo {year} {1993})}\BibitemShut {NoStop}%
\bibitem [{\citenamefont
  {Christodoulou}(1999{\natexlab{b}})}]{Christodoulou:1999aa}%
  \BibitemOpen
  \bibfield  {author} {\bibinfo {author} {\bibfnamefont {D.}~\bibnamefont
  {Christodoulou}},\ }\bibfield  {title} {\bibinfo {title} {The instability of
  naked singularities in the gravitational collapse of a scalar field},\
  }\href@noop {} {\bibfield  {journal} {\bibinfo  {journal} {Annals of
  Mathematics}\ }\textbf {\bibinfo {volume} {149}},\ \bibinfo {pages} {183}
  (\bibinfo {year} {1999}{\natexlab{b}})}\BibitemShut {NoStop}%
\bibitem [{\citenamefont {Christodoulou}(2009)}]{Christodoulou:2008nj}%
  \BibitemOpen
  \bibfield  {author} {\bibinfo {author} {\bibfnamefont {D.}~\bibnamefont
  {Christodoulou}},\ }\href {https://doi.org/10.4171/068} {\emph {\bibinfo
  {title} {{The Formation of Black Holes in General Relativity}}}},\ EMS
  Monographs in Mathematics\ (\bibinfo  {publisher} {European Mathematical
  Society Press},\ \bibinfo {year} {2009})\ \Eprint
  {https://arxiv.org/abs/0805.3880} {arXiv:0805.3880 [gr-qc]} \BibitemShut
  {NoStop}%
\bibitem [{\citenamefont {Whiting}(1989)}]{Whiting:1988vc}%
  \BibitemOpen
  \bibfield  {author} {\bibinfo {author} {\bibfnamefont {B.~F.}\ \bibnamefont
  {Whiting}},\ }\bibfield  {title} {\bibinfo {title} {{Mode Stability of the
  Kerr Black Hole}},\ }\href {https://doi.org/10.1063/1.528308} {\bibfield
  {journal} {\bibinfo  {journal} {J. Math. Phys.}\ }\textbf {\bibinfo {volume}
  {30}},\ \bibinfo {pages} {1301} (\bibinfo {year} {1989})}\BibitemShut
  {NoStop}%
\bibitem [{\citenamefont {Dafermos}\ and\ \citenamefont
  {Rodnianski}(2010)}]{Dafermos:2010hb}%
  \BibitemOpen
  \bibfield  {author} {\bibinfo {author} {\bibfnamefont {M.}~\bibnamefont
  {Dafermos}}\ and\ \bibinfo {author} {\bibfnamefont {I.}~\bibnamefont
  {Rodnianski}},\ }\bibfield  {title} {\bibinfo {title} {{Decay for solutions
  of the wave equation on Kerr exterior spacetimes I-II: The cases |a|
  \ensuremath{<}\ensuremath{<} M or axisymmetry}},\ }\Eprint
  {https://arxiv.org/abs/1010.5132} {arXiv:1010.5132 [gr-qc]}  (\bibinfo {year}
  {2010})\BibitemShut {NoStop}%
\bibitem [{\citenamefont {Dafermos}\ \emph {et~al.}(2014)\citenamefont
  {Dafermos}, \citenamefont {Rodnianski},\ and\ \citenamefont
  {Shlapentokh-Rothman}}]{Dafermos:2014cua}%
  \BibitemOpen
  \bibfield  {author} {\bibinfo {author} {\bibfnamefont {M.}~\bibnamefont
  {Dafermos}}, \bibinfo {author} {\bibfnamefont {I.}~\bibnamefont
  {Rodnianski}},\ and\ \bibinfo {author} {\bibfnamefont {Y.}~\bibnamefont
  {Shlapentokh-Rothman}},\ }\bibfield  {title} {\bibinfo {title} {{Decay for
  solutions of the wave equation on Kerr exterior spacetimes III: The full
  subextremal case |a| \ensuremath{<} M}},\ }\Eprint
  {https://arxiv.org/abs/1402.7034} {arXiv:1402.7034 [gr-qc]}  (\bibinfo {year}
  {2014})\BibitemShut {NoStop}%
\bibitem [{\citenamefont {Dafermos}\ \emph {et~al.}(2021)\citenamefont
  {Dafermos}, \citenamefont {Holzegel}, \citenamefont {Rodnianski},\ and\
  \citenamefont {Taylor}}]{Dafermos:2021cbw}%
  \BibitemOpen
  \bibfield  {author} {\bibinfo {author} {\bibfnamefont {M.}~\bibnamefont
  {Dafermos}}, \bibinfo {author} {\bibfnamefont {G.}~\bibnamefont {Holzegel}},
  \bibinfo {author} {\bibfnamefont {I.}~\bibnamefont {Rodnianski}},\ and\
  \bibinfo {author} {\bibfnamefont {M.}~\bibnamefont {Taylor}},\ }\bibfield
  {title} {\bibinfo {title} {{The non-linear stability of the Schwarzschild
  family of black holes}},\ }\Eprint {https://arxiv.org/abs/2104.08222}
  {arXiv:2104.08222 [gr-qc]}  (\bibinfo {year} {2021})\BibitemShut {NoStop}%
\bibitem [{\citenamefont {Klainerman}\ and\ \citenamefont
  {Szeftel}(2021)}]{Klainerman:2021qzy}%
  \BibitemOpen
  \bibfield  {author} {\bibinfo {author} {\bibfnamefont {S.}~\bibnamefont
  {Klainerman}}\ and\ \bibinfo {author} {\bibfnamefont {J.}~\bibnamefont
  {Szeftel}},\ }\bibfield  {title} {\bibinfo {title} {{Kerr stability for small
  angular momentum}},\ }\Eprint {https://arxiv.org/abs/2104.11857}
  {arXiv:2104.11857 [math.AP]}  (\bibinfo {year} {2021})\BibitemShut {NoStop}%
\bibitem [{\citenamefont {Gregory}\ and\ \citenamefont
  {Laflamme}(1993)}]{Gregory:1993vy}%
  \BibitemOpen
  \bibfield  {author} {\bibinfo {author} {\bibfnamefont {R.}~\bibnamefont
  {Gregory}}\ and\ \bibinfo {author} {\bibfnamefont {R.}~\bibnamefont
  {Laflamme}},\ }\bibfield  {title} {\bibinfo {title} {{Black strings and
  p-branes are unstable}},\ }\href
  {https://doi.org/10.1103/PhysRevLett.70.2837} {\bibfield  {journal} {\bibinfo
   {journal} {Phys. Rev. Lett.}\ }\textbf {\bibinfo {volume} {70}},\ \bibinfo
  {pages} {2837} (\bibinfo {year} {1993})},\ \Eprint
  {https://arxiv.org/abs/hep-th/9301052} {arXiv:hep-th/9301052} \BibitemShut
  {NoStop}%
\bibitem [{\citenamefont {Emparan}\ and\ \citenamefont
  {Reall}(2002)}]{Emparan:2001wn}%
  \BibitemOpen
  \bibfield  {author} {\bibinfo {author} {\bibfnamefont {R.}~\bibnamefont
  {Emparan}}\ and\ \bibinfo {author} {\bibfnamefont {H.~S.}\ \bibnamefont
  {Reall}},\ }\bibfield  {title} {\bibinfo {title} {{A Rotating black ring
  solution in five-dimensions}},\ }\href
  {https://doi.org/10.1103/PhysRevLett.88.101101} {\bibfield  {journal}
  {\bibinfo  {journal} {Phys. Rev. Lett.}\ }\textbf {\bibinfo {volume} {88}},\
  \bibinfo {pages} {101101} (\bibinfo {year} {2002})},\ \Eprint
  {https://arxiv.org/abs/hep-th/0110260} {arXiv:hep-th/0110260} \BibitemShut
  {NoStop}%
\bibitem [{\citenamefont {Emparan}\ and\ \citenamefont
  {Reall}(2008)}]{Emparan:2008eg}%
  \BibitemOpen
  \bibfield  {author} {\bibinfo {author} {\bibfnamefont {R.}~\bibnamefont
  {Emparan}}\ and\ \bibinfo {author} {\bibfnamefont {H.~S.}\ \bibnamefont
  {Reall}},\ }\bibfield  {title} {\bibinfo {title} {{Black Holes in Higher
  Dimensions}},\ }\href {https://doi.org/10.12942/lrr-2008-6} {\bibfield
  {journal} {\bibinfo  {journal} {Living Rev. Rel.}\ }\textbf {\bibinfo
  {volume} {11}},\ \bibinfo {pages} {6} (\bibinfo {year} {2008})},\ \Eprint
  {https://arxiv.org/abs/0801.3471} {arXiv:0801.3471 [hep-th]} \BibitemShut
  {NoStop}%
\bibitem [{\citenamefont {Emparan}\ and\ \citenamefont
  {Myers}(2003)}]{Emparan:2003sy}%
  \BibitemOpen
  \bibfield  {author} {\bibinfo {author} {\bibfnamefont {R.}~\bibnamefont
  {Emparan}}\ and\ \bibinfo {author} {\bibfnamefont {R.~C.}\ \bibnamefont
  {Myers}},\ }\bibfield  {title} {\bibinfo {title} {{Instability of
  ultra-spinning black holes}},\ }\href
  {https://doi.org/10.1088/1126-6708/2003/09/025} {\bibfield  {journal}
  {\bibinfo  {journal} {JHEP}\ }\textbf {\bibinfo {volume} {09}},\ \bibinfo
  {pages} {025}},\ \Eprint {https://arxiv.org/abs/hep-th/0308056}
  {arXiv:hep-th/0308056} \BibitemShut {NoStop}%
\bibitem [{\citenamefont {Dias}\ \emph {et~al.}(2009)\citenamefont {Dias},
  \citenamefont {Figueras}, \citenamefont {Monteiro}, \citenamefont {Santos},\
  and\ \citenamefont {Emparan}}]{Dias:2009iu}%
  \BibitemOpen
  \bibfield  {author} {\bibinfo {author} {\bibfnamefont {O.~J.~C.}\
  \bibnamefont {Dias}}, \bibinfo {author} {\bibfnamefont {P.}~\bibnamefont
  {Figueras}}, \bibinfo {author} {\bibfnamefont {R.}~\bibnamefont {Monteiro}},
  \bibinfo {author} {\bibfnamefont {J.~E.}\ \bibnamefont {Santos}},\ and\
  \bibinfo {author} {\bibfnamefont {R.}~\bibnamefont {Emparan}},\ }\bibfield
  {title} {\bibinfo {title} {{Instability and new phases of higher-dimensional
  rotating black holes}},\ }\href {https://doi.org/10.1103/PhysRevD.80.111701}
  {\bibfield  {journal} {\bibinfo  {journal} {Phys. Rev. D}\ }\textbf {\bibinfo
  {volume} {80}},\ \bibinfo {pages} {111701} (\bibinfo {year} {2009})},\
  \Eprint {https://arxiv.org/abs/0907.2248} {arXiv:0907.2248 [hep-th]}
  \BibitemShut {NoStop}%
\bibitem [{\citenamefont {Dias}\ \emph {et~al.}(2010)\citenamefont {Dias},
  \citenamefont {Figueras}, \citenamefont {Monteiro}, \citenamefont {Reall},\
  and\ \citenamefont {Santos}}]{Dias:2010eu}%
  \BibitemOpen
  \bibfield  {author} {\bibinfo {author} {\bibfnamefont {O.~J.~C.}\
  \bibnamefont {Dias}}, \bibinfo {author} {\bibfnamefont {P.}~\bibnamefont
  {Figueras}}, \bibinfo {author} {\bibfnamefont {R.}~\bibnamefont {Monteiro}},
  \bibinfo {author} {\bibfnamefont {H.~S.}\ \bibnamefont {Reall}},\ and\
  \bibinfo {author} {\bibfnamefont {J.~E.}\ \bibnamefont {Santos}},\ }\bibfield
   {title} {\bibinfo {title} {{An instability of higher-dimensional rotating
  black holes}},\ }\href {https://doi.org/10.1007/JHEP05(2010)076} {\bibfield
  {journal} {\bibinfo  {journal} {JHEP}\ }\textbf {\bibinfo {volume} {05}},\
  \bibinfo {pages} {076}},\ \Eprint {https://arxiv.org/abs/1001.4527}
  {arXiv:1001.4527 [hep-th]} \BibitemShut {NoStop}%
\bibitem [{\citenamefont {Dias}\ \emph {et~al.}(2014)\citenamefont {Dias},
  \citenamefont {Hartnett},\ and\ \citenamefont {Santos}}]{Dias:2014eua}%
  \BibitemOpen
  \bibfield  {author} {\bibinfo {author} {\bibfnamefont {O.~J.~C.}\
  \bibnamefont {Dias}}, \bibinfo {author} {\bibfnamefont {G.~S.}\ \bibnamefont
  {Hartnett}},\ and\ \bibinfo {author} {\bibfnamefont {J.~E.}\ \bibnamefont
  {Santos}},\ }\bibfield  {title} {\bibinfo {title} {{Quasinormal modes of
  asymptotically flat rotating black holes}},\ }\href
  {https://doi.org/10.1088/0264-9381/31/24/245011} {\bibfield  {journal}
  {\bibinfo  {journal} {Class. Quant. Grav.}\ }\textbf {\bibinfo {volume}
  {31}},\ \bibinfo {pages} {245011} (\bibinfo {year} {2014})},\ \Eprint
  {https://arxiv.org/abs/1402.7047} {arXiv:1402.7047 [hep-th]} \BibitemShut
  {NoStop}%
\bibitem [{\citenamefont {Santos}\ and\ \citenamefont
  {Way}(2015)}]{Santos:2015iua}%
  \BibitemOpen
  \bibfield  {author} {\bibinfo {author} {\bibfnamefont {J.~E.}\ \bibnamefont
  {Santos}}\ and\ \bibinfo {author} {\bibfnamefont {B.}~\bibnamefont {Way}},\
  }\bibfield  {title} {\bibinfo {title} {{Neutral Black Rings in Five
  Dimensions are Unstable}},\ }\href
  {https://doi.org/10.1103/PhysRevLett.114.221101} {\bibfield  {journal}
  {\bibinfo  {journal} {Phys. Rev. Lett.}\ }\textbf {\bibinfo {volume} {114}},\
  \bibinfo {pages} {221101} (\bibinfo {year} {2015})},\ \Eprint
  {https://arxiv.org/abs/1503.00721} {arXiv:1503.00721 [hep-th]} \BibitemShut
  {NoStop}%
\bibitem [{\citenamefont {Hubeny}\ and\ \citenamefont
  {Rangamani}(2002)}]{Hubeny:2002xn}%
  \BibitemOpen
  \bibfield  {author} {\bibinfo {author} {\bibfnamefont {V.~E.}\ \bibnamefont
  {Hubeny}}\ and\ \bibinfo {author} {\bibfnamefont {M.}~\bibnamefont
  {Rangamani}},\ }\bibfield  {title} {\bibinfo {title} {{Unstable horizons}},\
  }\href {https://doi.org/10.1088/1126-6708/2002/05/027} {\bibfield  {journal}
  {\bibinfo  {journal} {JHEP}\ }\textbf {\bibinfo {volume} {05}},\ \bibinfo
  {pages} {027}},\ \Eprint {https://arxiv.org/abs/hep-th/0202189}
  {arXiv:hep-th/0202189} \BibitemShut {NoStop}%
\bibitem [{\citenamefont {Hirayama}\ and\ \citenamefont
  {Kang}(2001)}]{Hirayama:2001bi}%
  \BibitemOpen
  \bibfield  {author} {\bibinfo {author} {\bibfnamefont {T.}~\bibnamefont
  {Hirayama}}\ and\ \bibinfo {author} {\bibfnamefont {G.}~\bibnamefont
  {Kang}},\ }\bibfield  {title} {\bibinfo {title} {{Stable black strings in
  anti-de Sitter space}},\ }\href {https://doi.org/10.1103/PhysRevD.64.064010}
  {\bibfield  {journal} {\bibinfo  {journal} {Phys. Rev. D}\ }\textbf {\bibinfo
  {volume} {64}},\ \bibinfo {pages} {064010} (\bibinfo {year} {2001})},\
  \Eprint {https://arxiv.org/abs/hep-th/0104213} {arXiv:hep-th/0104213}
  \BibitemShut {NoStop}%
\bibitem [{\citenamefont {Lehner}\ and\ \citenamefont
  {Pretorius}(2010)}]{Lehner:2010pn}%
  \BibitemOpen
  \bibfield  {author} {\bibinfo {author} {\bibfnamefont {L.}~\bibnamefont
  {Lehner}}\ and\ \bibinfo {author} {\bibfnamefont {F.}~\bibnamefont
  {Pretorius}},\ }\bibfield  {title} {\bibinfo {title} {{Black Strings, Low
  Viscosity Fluids, and Violation of Cosmic Censorship}},\ }\href
  {https://doi.org/10.1103/PhysRevLett.105.101102} {\bibfield  {journal}
  {\bibinfo  {journal} {Phys. Rev. Lett.}\ }\textbf {\bibinfo {volume} {105}},\
  \bibinfo {pages} {101102} (\bibinfo {year} {2010})},\ \Eprint
  {https://arxiv.org/abs/1006.5960} {arXiv:1006.5960 [hep-th]} \BibitemShut
  {NoStop}%
\bibitem [{\citenamefont {Lehner}\ and\ \citenamefont
  {Pretorius}(2011)}]{Lehner:2011wc}%
  \BibitemOpen
  \bibfield  {author} {\bibinfo {author} {\bibfnamefont {L.}~\bibnamefont
  {Lehner}}\ and\ \bibinfo {author} {\bibfnamefont {F.}~\bibnamefont
  {Pretorius}},\ }\bibfield  {title} {\bibinfo {title} {{Final State of
  Gregory-Laflamme Instability}},\ }\Eprint {https://arxiv.org/abs/1106.5184}
  {arXiv:1106.5184 [gr-qc]}  (\bibinfo {year} {2011})\BibitemShut {NoStop}%
\bibitem [{\citenamefont {Hawking}\ and\ \citenamefont
  {Ellis}(2011)}]{Hawking:1973uf}%
  \BibitemOpen
  \bibfield  {author} {\bibinfo {author} {\bibfnamefont {S.~W.}\ \bibnamefont
  {Hawking}}\ and\ \bibinfo {author} {\bibfnamefont {G.~F.~R.}\ \bibnamefont
  {Ellis}},\ }\href {https://doi.org/10.1017/CBO9780511524646} {\emph {\bibinfo
  {title} {{The Large Scale Structure of Space-Time}}}},\ Cambridge Monographs
  on Mathematical Physics\ (\bibinfo  {publisher} {Cambridge University
  Press},\ \bibinfo {year} {2011})\BibitemShut {NoStop}%
\bibitem [{\citenamefont {Figueras}\ \emph {et~al.}(2016)\citenamefont
  {Figueras}, \citenamefont {Kunesch},\ and\ \citenamefont
  {Tunyasuvunakool}}]{Figueras:2015hkb}%
  \BibitemOpen
  \bibfield  {author} {\bibinfo {author} {\bibfnamefont {P.}~\bibnamefont
  {Figueras}}, \bibinfo {author} {\bibfnamefont {M.}~\bibnamefont {Kunesch}},\
  and\ \bibinfo {author} {\bibfnamefont {S.}~\bibnamefont {Tunyasuvunakool}},\
  }\bibfield  {title} {\bibinfo {title} {{End Point of Black Ring Instabilities
  and the Weak Cosmic Censorship Conjecture}},\ }\href
  {https://doi.org/10.1103/PhysRevLett.116.071102} {\bibfield  {journal}
  {\bibinfo  {journal} {Phys. Rev. Lett.}\ }\textbf {\bibinfo {volume} {116}},\
  \bibinfo {pages} {071102} (\bibinfo {year} {2016})},\ \Eprint
  {https://arxiv.org/abs/1512.04532} {arXiv:1512.04532 [hep-th]} \BibitemShut
  {NoStop}%
\bibitem [{\citenamefont {Figueras}\ \emph {et~al.}(2017)\citenamefont
  {Figueras}, \citenamefont {Kunesch}, \citenamefont {Lehner},\ and\
  \citenamefont {Tunyasuvunakool}}]{Figueras:2017zwa}%
  \BibitemOpen
  \bibfield  {author} {\bibinfo {author} {\bibfnamefont {P.}~\bibnamefont
  {Figueras}}, \bibinfo {author} {\bibfnamefont {M.}~\bibnamefont {Kunesch}},
  \bibinfo {author} {\bibfnamefont {L.}~\bibnamefont {Lehner}},\ and\ \bibinfo
  {author} {\bibfnamefont {S.}~\bibnamefont {Tunyasuvunakool}},\ }\bibfield
  {title} {\bibinfo {title} {{End Point of the Ultraspinning Instability and
  Violation of Cosmic Censorship}},\ }\href
  {https://doi.org/10.1103/PhysRevLett.118.151103} {\bibfield  {journal}
  {\bibinfo  {journal} {Phys. Rev. Lett.}\ }\textbf {\bibinfo {volume} {118}},\
  \bibinfo {pages} {151103} (\bibinfo {year} {2017})},\ \Eprint
  {https://arxiv.org/abs/1702.01755} {arXiv:1702.01755 [hep-th]} \BibitemShut
  {NoStop}%
\bibitem [{\citenamefont {Bantilan}\ \emph {et~al.}(2019)\citenamefont
  {Bantilan}, \citenamefont {Figueras}, \citenamefont {Kunesch},\ and\
  \citenamefont {Panosso~Macedo}}]{Bantilan:2019bvf}%
  \BibitemOpen
  \bibfield  {author} {\bibinfo {author} {\bibfnamefont {H.}~\bibnamefont
  {Bantilan}}, \bibinfo {author} {\bibfnamefont {P.}~\bibnamefont {Figueras}},
  \bibinfo {author} {\bibfnamefont {M.}~\bibnamefont {Kunesch}},\ and\ \bibinfo
  {author} {\bibfnamefont {R.}~\bibnamefont {Panosso~Macedo}},\ }\bibfield
  {title} {\bibinfo {title} {{End point of nonaxisymmetric black hole
  instabilities in higher dimensions}},\ }\href
  {https://doi.org/10.1103/PhysRevD.100.086014} {\bibfield  {journal} {\bibinfo
   {journal} {Phys. Rev. D}\ }\textbf {\bibinfo {volume} {100}},\ \bibinfo
  {pages} {086014} (\bibinfo {year} {2019})},\ \Eprint
  {https://arxiv.org/abs/1906.10696} {arXiv:1906.10696 [hep-th]} \BibitemShut
  {NoStop}%
\bibitem [{\citenamefont {Andrade}\ \emph {et~al.}(2022)\citenamefont
  {Andrade}, \citenamefont {Figueras},\ and\ \citenamefont
  {Sperhake}}]{Andrade:2020dgc}%
  \BibitemOpen
  \bibfield  {author} {\bibinfo {author} {\bibfnamefont {T.}~\bibnamefont
  {Andrade}}, \bibinfo {author} {\bibfnamefont {P.}~\bibnamefont {Figueras}},\
  and\ \bibinfo {author} {\bibfnamefont {U.}~\bibnamefont {Sperhake}},\
  }\bibfield  {title} {\bibinfo {title} {{Evidence for violations of Weak
  Cosmic Censorship in black hole collisions in higher dimensions}},\ }\href
  {https://doi.org/10.1007/JHEP03(2022)111} {\bibfield  {journal} {\bibinfo
  {journal} {JHEP}\ }\textbf {\bibinfo {volume} {03}},\ \bibinfo {pages}
  {111}},\ \Eprint {https://arxiv.org/abs/2011.03049} {arXiv:2011.03049
  [hep-th]} \BibitemShut {NoStop}%
\bibitem [{\citenamefont {Andrade}\ \emph
  {et~al.}(2019{\natexlab{a}})\citenamefont {Andrade}, \citenamefont {Emparan},
  \citenamefont {Licht},\ and\ \citenamefont {Luna}}]{Andrade:2018yqu}%
  \BibitemOpen
  \bibfield  {author} {\bibinfo {author} {\bibfnamefont {T.}~\bibnamefont
  {Andrade}}, \bibinfo {author} {\bibfnamefont {R.}~\bibnamefont {Emparan}},
  \bibinfo {author} {\bibfnamefont {D.}~\bibnamefont {Licht}},\ and\ \bibinfo
  {author} {\bibfnamefont {R.}~\bibnamefont {Luna}},\ }\bibfield  {title}
  {\bibinfo {title} {{Cosmic censorship violation in black hole collisions in
  higher dimensions}},\ }\href {https://doi.org/10.1007/JHEP04(2019)121}
  {\bibfield  {journal} {\bibinfo  {journal} {JHEP}\ }\textbf {\bibinfo
  {volume} {04}},\ \bibinfo {pages} {121}},\ \Eprint
  {https://arxiv.org/abs/1812.05017} {arXiv:1812.05017 [hep-th]} \BibitemShut
  {NoStop}%
\bibitem [{\citenamefont {Andrade}\ \emph
  {et~al.}(2019{\natexlab{b}})\citenamefont {Andrade}, \citenamefont {Emparan},
  \citenamefont {Licht},\ and\ \citenamefont {Luna}}]{Andrade:2019edf}%
  \BibitemOpen
  \bibfield  {author} {\bibinfo {author} {\bibfnamefont {T.}~\bibnamefont
  {Andrade}}, \bibinfo {author} {\bibfnamefont {R.}~\bibnamefont {Emparan}},
  \bibinfo {author} {\bibfnamefont {D.}~\bibnamefont {Licht}},\ and\ \bibinfo
  {author} {\bibfnamefont {R.}~\bibnamefont {Luna}},\ }\bibfield  {title}
  {\bibinfo {title} {{Black hole collisions, instabilities, and cosmic
  censorship violation at large $D$}},\ }\href
  {https://doi.org/10.1007/JHEP09(2019)099} {\bibfield  {journal} {\bibinfo
  {journal} {JHEP}\ }\textbf {\bibinfo {volume} {09}},\ \bibinfo {pages}
  {099}},\ \Eprint {https://arxiv.org/abs/1908.03424} {arXiv:1908.03424
  [hep-th]} \BibitemShut {NoStop}%
\bibitem [{\citenamefont {Andrade}\ \emph {et~al.}(2020)\citenamefont
  {Andrade}, \citenamefont {Emparan}, \citenamefont {Jansen}, \citenamefont
  {Licht}, \citenamefont {Luna},\ and\ \citenamefont
  {Suzuki}}]{Andrade:2020ilm}%
  \BibitemOpen
  \bibfield  {author} {\bibinfo {author} {\bibfnamefont {T.}~\bibnamefont
  {Andrade}}, \bibinfo {author} {\bibfnamefont {R.}~\bibnamefont {Emparan}},
  \bibinfo {author} {\bibfnamefont {A.}~\bibnamefont {Jansen}}, \bibinfo
  {author} {\bibfnamefont {D.}~\bibnamefont {Licht}}, \bibinfo {author}
  {\bibfnamefont {R.}~\bibnamefont {Luna}},\ and\ \bibinfo {author}
  {\bibfnamefont {R.}~\bibnamefont {Suzuki}},\ }\bibfield  {title} {\bibinfo
  {title} {{Entropy production and entropic attractors in black hole fusion and
  fission}},\ }\href {https://doi.org/10.1007/JHEP08(2020)098} {\bibfield
  {journal} {\bibinfo  {journal} {JHEP}\ }\textbf {\bibinfo {volume} {08}},\
  \bibinfo {pages} {098}},\ \Eprint {https://arxiv.org/abs/2005.14498}
  {arXiv:2005.14498 [hep-th]} \BibitemShut {NoStop}%
\bibitem [{\citenamefont {Emparan}\ \emph {et~al.}(2022)\citenamefont
  {Emparan}, \citenamefont {Licht}, \citenamefont {Suzuki}, \citenamefont
  {Toma\v{s}evi\'c},\ and\ \citenamefont {Way}}]{Emparan:2021ewh}%
  \BibitemOpen
  \bibfield  {author} {\bibinfo {author} {\bibfnamefont {R.}~\bibnamefont
  {Emparan}}, \bibinfo {author} {\bibfnamefont {D.}~\bibnamefont {Licht}},
  \bibinfo {author} {\bibfnamefont {R.}~\bibnamefont {Suzuki}}, \bibinfo
  {author} {\bibfnamefont {M.}~\bibnamefont {Toma\v{s}evi\'c}},\ and\ \bibinfo
  {author} {\bibfnamefont {B.}~\bibnamefont {Way}},\ }\bibfield  {title}
  {\bibinfo {title} {{Black tsunamis and naked singularities in AdS}},\ }\href
  {https://doi.org/10.1007/JHEP02(2022)090} {\bibfield  {journal} {\bibinfo
  {journal} {JHEP}\ }\textbf {\bibinfo {volume} {02}},\ \bibinfo {pages}
  {090}},\ \Eprint {https://arxiv.org/abs/2112.07967} {arXiv:2112.07967
  [hep-th]} \BibitemShut {NoStop}%
\bibitem [{\citenamefont {Alic}\ \emph {et~al.}(2012)\citenamefont {Alic},
  \citenamefont {Bona-Casas}, \citenamefont {Bona}, \citenamefont {Rezzolla},\
  and\ \citenamefont {Palenzuela}}]{Alic:2011gg}%
  \BibitemOpen
  \bibfield  {author} {\bibinfo {author} {\bibfnamefont {D.}~\bibnamefont
  {Alic}}, \bibinfo {author} {\bibfnamefont {C.}~\bibnamefont {Bona-Casas}},
  \bibinfo {author} {\bibfnamefont {C.}~\bibnamefont {Bona}}, \bibinfo {author}
  {\bibfnamefont {L.}~\bibnamefont {Rezzolla}},\ and\ \bibinfo {author}
  {\bibfnamefont {C.}~\bibnamefont {Palenzuela}},\ }\bibfield  {title}
  {\bibinfo {title} {{Conformal and covariant formulation of the Z4 system with
  constraint-violation damping}},\ }\href
  {https://doi.org/10.1103/PhysRevD.85.064040} {\bibfield  {journal} {\bibinfo
  {journal} {Phys. Rev. D}\ }\textbf {\bibinfo {volume} {85}},\ \bibinfo
  {pages} {064040} (\bibinfo {year} {2012})},\ \Eprint
  {https://arxiv.org/abs/1106.2254} {arXiv:1106.2254 [gr-qc]} \BibitemShut
  {NoStop}%
\bibitem [{\citenamefont {Alic}\ \emph {et~al.}(2013)\citenamefont {Alic},
  \citenamefont {Kastaun},\ and\ \citenamefont {Rezzolla}}]{Alic:2013xsa}%
  \BibitemOpen
  \bibfield  {author} {\bibinfo {author} {\bibfnamefont {D.}~\bibnamefont
  {Alic}}, \bibinfo {author} {\bibfnamefont {W.}~\bibnamefont {Kastaun}},\ and\
  \bibinfo {author} {\bibfnamefont {L.}~\bibnamefont {Rezzolla}},\ }\bibfield
  {title} {\bibinfo {title} {{Constraint damping of the conformal and covariant
  formulation of the Z4 system in simulations of binary neutron stars}},\
  }\href {https://doi.org/10.1103/PhysRevD.88.064049} {\bibfield  {journal}
  {\bibinfo  {journal} {Phys. Rev. D}\ }\textbf {\bibinfo {volume} {88}},\
  \bibinfo {pages} {064049} (\bibinfo {year} {2013})},\ \Eprint
  {https://arxiv.org/abs/1307.7391} {arXiv:1307.7391 [gr-qc]} \BibitemShut
  {NoStop}%
\bibitem [{\citenamefont {Clough}\ \emph {et~al.}(2015)\citenamefont {Clough},
  \citenamefont {Figueras}, \citenamefont {Finkel}, \citenamefont {Kunesch},
  \citenamefont {Lim},\ and\ \citenamefont {Tunyasuvunakool}}]{Clough:2015sqa}%
  \BibitemOpen
  \bibfield  {author} {\bibinfo {author} {\bibfnamefont {K.}~\bibnamefont
  {Clough}}, \bibinfo {author} {\bibfnamefont {P.}~\bibnamefont {Figueras}},
  \bibinfo {author} {\bibfnamefont {H.}~\bibnamefont {Finkel}}, \bibinfo
  {author} {\bibfnamefont {M.}~\bibnamefont {Kunesch}}, \bibinfo {author}
  {\bibfnamefont {E.~A.}\ \bibnamefont {Lim}},\ and\ \bibinfo {author}
  {\bibfnamefont {S.}~\bibnamefont {Tunyasuvunakool}},\ }\bibfield  {title}
  {\bibinfo {title} {{GRChombo : Numerical Relativity with Adaptive Mesh
  Refinement}},\ }\href {https://doi.org/10.1088/0264-9381/32/24/245011}
  {\bibfield  {journal} {\bibinfo  {journal} {Class. Quant. Grav.}\ }\textbf
  {\bibinfo {volume} {32}},\ \bibinfo {pages} {245011} (\bibinfo {year}
  {2015})},\ \Eprint {https://arxiv.org/abs/1503.03436} {arXiv:1503.03436
  [gr-qc]} \BibitemShut {NoStop}%
\bibitem [{\citenamefont {Andrade}\ \emph {et~al.}(2021)\citenamefont {Andrade}
  \emph {et~al.}}]{Andrade:2021rbd}%
  \BibitemOpen
  \bibfield  {author} {\bibinfo {author} {\bibfnamefont {T.}~\bibnamefont
  {Andrade}} \emph {et~al.},\ }\bibfield  {title} {\bibinfo {title} {{GRChombo:
  An adaptable numerical relativity code for fundamental physics}},\ }\href
  {https://doi.org/10.21105/joss.03703} {\bibfield  {journal} {\bibinfo
  {journal} {J. Open Source Softw.}\ }\textbf {\bibinfo {volume} {6}},\
  \bibinfo {pages} {3703} (\bibinfo {year} {2021})},\ \Eprint
  {https://arxiv.org/abs/2201.03458} {arXiv:2201.03458 [gr-qc]} \BibitemShut
  {NoStop}%
\bibitem [{\citenamefont {Eggers}(1993)}]{Eggers:1993aa}%
  \BibitemOpen
  \bibfield  {author} {\bibinfo {author} {\bibfnamefont {J.}~\bibnamefont
  {Eggers}},\ }\bibfield  {title} {\bibinfo {title} {Universal pinching of 3d
  axisymmetric free-surface flow},\ }\href@noop {} {\bibfield  {journal}
  {\bibinfo  {journal} {Phys. Rev. Lett.}\ }\textbf {\bibinfo {volume} {71}},\
  \bibinfo {pages} {3458} (\bibinfo {year} {1993})}\BibitemShut {NoStop}%
\bibitem [{\citenamefont {Eggers}(1997)}]{Eggers:1997aa}%
  \BibitemOpen
  \bibfield  {author} {\bibinfo {author} {\bibfnamefont {J.}~\bibnamefont
  {Eggers}},\ }\bibfield  {title} {\bibinfo {title} {Nonlinear dynamics and
  breakup of free-surface flows},\ }\href@noop {} {\bibfield  {journal}
  {\bibinfo  {journal} {Rev. Mod. Phys.}\ }\textbf {\bibinfo {volume} {69}},\
  \bibinfo {pages} {865} (\bibinfo {year} {1997})}\BibitemShut {NoStop}%
\bibitem [{\citenamefont {Pretorius}(2005)}]{Pretorius:2004jg}%
  \BibitemOpen
  \bibfield  {author} {\bibinfo {author} {\bibfnamefont {F.}~\bibnamefont
  {Pretorius}},\ }\bibfield  {title} {\bibinfo {title} {{Numerical relativity
  using a generalized harmonic decomposition}},\ }\href
  {https://doi.org/10.1088/0264-9381/22/2/014} {\bibfield  {journal} {\bibinfo
  {journal} {Class. Quant. Grav.}\ }\textbf {\bibinfo {volume} {22}},\ \bibinfo
  {pages} {425} (\bibinfo {year} {2005})},\ \Eprint
  {https://arxiv.org/abs/gr-qc/0407110} {arXiv:gr-qc/0407110} \BibitemShut
  {NoStop}%
\bibitem [{\citenamefont {Shibata}\ and\ \citenamefont
  {Yoshino}(2010{\natexlab{a}})}]{Shibata:2010wz}%
  \BibitemOpen
  \bibfield  {author} {\bibinfo {author} {\bibfnamefont {M.}~\bibnamefont
  {Shibata}}\ and\ \bibinfo {author} {\bibfnamefont {H.}~\bibnamefont
  {Yoshino}},\ }\bibfield  {title} {\bibinfo {title} {{Bar-mode instability of
  rapidly spinning black hole in higher dimensions: Numerical simulation in
  general relativity}},\ }\href {https://doi.org/10.1103/PhysRevD.81.104035}
  {\bibfield  {journal} {\bibinfo  {journal} {Phys. Rev. D}\ }\textbf {\bibinfo
  {volume} {81}},\ \bibinfo {pages} {104035} (\bibinfo {year}
  {2010}{\natexlab{a}})},\ \Eprint {https://arxiv.org/abs/1004.4970}
  {arXiv:1004.4970 [gr-qc]} \BibitemShut {NoStop}%
\bibitem [{\citenamefont {Cook}\ \emph {et~al.}(2016)\citenamefont {Cook},
  \citenamefont {Figueras}, \citenamefont {Kunesch}, \citenamefont {Sperhake},\
  and\ \citenamefont {Tunyasuvunakool}}]{Cook:2016soy}%
  \BibitemOpen
  \bibfield  {author} {\bibinfo {author} {\bibfnamefont {W.~G.}\ \bibnamefont
  {Cook}}, \bibinfo {author} {\bibfnamefont {P.}~\bibnamefont {Figueras}},
  \bibinfo {author} {\bibfnamefont {M.}~\bibnamefont {Kunesch}}, \bibinfo
  {author} {\bibfnamefont {U.}~\bibnamefont {Sperhake}},\ and\ \bibinfo
  {author} {\bibfnamefont {S.}~\bibnamefont {Tunyasuvunakool}},\ }\bibfield
  {title} {\bibinfo {title} {{Dimensional reduction in numerical relativity:
  Modified cartoon formalism and regularization}},\ }\href
  {https://doi.org/10.1142/S0218271816410133} {\bibfield  {journal} {\bibinfo
  {journal} {Int. J. Mod. Phys. D}\ }\textbf {\bibinfo {volume} {25}},\
  \bibinfo {pages} {1641013} (\bibinfo {year} {2016})},\ \Eprint
  {https://arxiv.org/abs/1603.00362} {arXiv:1603.00362 [gr-qc]} \BibitemShut
  {NoStop}%
\bibitem [{\citenamefont {Shibata}\ and\ \citenamefont
  {Yoshino}(2010{\natexlab{b}})}]{Shibata:2009ad}%
  \BibitemOpen
  \bibfield  {author} {\bibinfo {author} {\bibfnamefont {M.}~\bibnamefont
  {Shibata}}\ and\ \bibinfo {author} {\bibfnamefont {H.}~\bibnamefont
  {Yoshino}},\ }\bibfield  {title} {\bibinfo {title} {{Nonaxisymmetric
  instability of rapidly rotating black hole in five dimensions}},\ }\href
  {https://doi.org/10.1103/PhysRevD.81.021501} {\bibfield  {journal} {\bibinfo
  {journal} {Phys. Rev. D}\ }\textbf {\bibinfo {volume} {81}},\ \bibinfo
  {pages} {021501} (\bibinfo {year} {2010}{\natexlab{b}})},\ \Eprint
  {https://arxiv.org/abs/0912.3606} {arXiv:0912.3606 [gr-qc]} \BibitemShut
  {NoStop}%
\bibitem [{\citenamefont {Painlev\'e}(1921)}]{Painleve}%
  \BibitemOpen
  \bibfield  {author} {\bibinfo {author} {\bibfnamefont {P.}~\bibnamefont
  {Painlev\'e}},\ }\bibfield  {title} {\bibinfo {title} {{La m\'ecanique
  classique et la th\'eorie de la relativit\'e}},\ }\href@noop {} {\bibfield
  {journal} {\bibinfo  {journal} {C. R. Acad. Sci. (Paris)}\ }\textbf {\bibinfo
  {volume} {173}},\ \bibinfo {pages} {677} (\bibinfo {year}
  {1921})}\BibitemShut {NoStop}%
\bibitem [{\citenamefont {Gullstrand}(1922)}]{Gullstrand}%
  \BibitemOpen
  \bibfield  {author} {\bibinfo {author} {\bibfnamefont {A.}~\bibnamefont
  {Gullstrand}},\ }\bibfield  {title} {\bibinfo {title} {{Allgemeine L\"{o}sung
  des statischen Eink\"{o}rperproblems in der Einsteinschen
  Gravitationstheorie}},\ }\href@noop {} {\bibfield  {journal} {\bibinfo
  {journal} {Arkiv for Matematik, Astronomi och Fysik}\ }\textbf {\bibinfo
  {volume} {16}},\ \bibinfo {pages} {1} (\bibinfo {year} {1922})}\BibitemShut
  {NoStop}%
\bibitem [{\citenamefont {Brown}\ \emph {et~al.}(2007)\citenamefont {Brown},
  \citenamefont {Sarbach}, \citenamefont {Schnetter}, \citenamefont {Tiglio},
  \citenamefont {Diener}, \citenamefont {Hawke},\ and\ \citenamefont
  {Pollney}}]{Brown:2007pg}%
  \BibitemOpen
  \bibfield  {author} {\bibinfo {author} {\bibfnamefont {J.~D.}\ \bibnamefont
  {Brown}}, \bibinfo {author} {\bibfnamefont {O.}~\bibnamefont {Sarbach}},
  \bibinfo {author} {\bibfnamefont {E.}~\bibnamefont {Schnetter}}, \bibinfo
  {author} {\bibfnamefont {M.}~\bibnamefont {Tiglio}}, \bibinfo {author}
  {\bibfnamefont {P.}~\bibnamefont {Diener}}, \bibinfo {author} {\bibfnamefont
  {I.}~\bibnamefont {Hawke}},\ and\ \bibinfo {author} {\bibfnamefont
  {D.}~\bibnamefont {Pollney}},\ }\bibfield  {title} {\bibinfo {title}
  {{Excision without excision: The Relativistic turducken}},\ }\href
  {https://doi.org/10.1103/PhysRevD.76.081503} {\bibfield  {journal} {\bibinfo
  {journal} {Phys. Rev. D}\ }\textbf {\bibinfo {volume} {76}},\ \bibinfo
  {pages} {081503} (\bibinfo {year} {2007})},\ \Eprint
  {https://arxiv.org/abs/0707.3101} {arXiv:0707.3101 [gr-qc]} \BibitemShut
  {NoStop}%
\bibitem [{\citenamefont {Brown}\ \emph {et~al.}(2009)\citenamefont {Brown},
  \citenamefont {Diener}, \citenamefont {Sarbach}, \citenamefont {Schnetter},\
  and\ \citenamefont {Tiglio}}]{Brown:2008sb}%
  \BibitemOpen
  \bibfield  {author} {\bibinfo {author} {\bibfnamefont {J.~D.}\ \bibnamefont
  {Brown}}, \bibinfo {author} {\bibfnamefont {P.}~\bibnamefont {Diener}},
  \bibinfo {author} {\bibfnamefont {O.}~\bibnamefont {Sarbach}}, \bibinfo
  {author} {\bibfnamefont {E.}~\bibnamefont {Schnetter}},\ and\ \bibinfo
  {author} {\bibfnamefont {M.}~\bibnamefont {Tiglio}},\ }\bibfield  {title}
  {\bibinfo {title} {{Turduckening black holes: An Analytical and computational
  study}},\ }\href {https://doi.org/10.1103/PhysRevD.79.044023} {\bibfield
  {journal} {\bibinfo  {journal} {Phys. Rev. D}\ }\textbf {\bibinfo {volume}
  {79}},\ \bibinfo {pages} {044023} (\bibinfo {year} {2009})},\ \Eprint
  {https://arxiv.org/abs/0809.3533} {arXiv:0809.3533 [gr-qc]} \BibitemShut
  {NoStop}%
\bibitem [{\citenamefont {Balay}\ \emph {et~al.}(1997)\citenamefont {Balay},
  \citenamefont {Gropp}, \citenamefont {McInnes},\ and\ \citenamefont
  {Smith}}]{petsc-efficient}%
  \BibitemOpen
  \bibfield  {author} {\bibinfo {author} {\bibfnamefont {S.}~\bibnamefont
  {Balay}}, \bibinfo {author} {\bibfnamefont {W.~D.}\ \bibnamefont {Gropp}},
  \bibinfo {author} {\bibfnamefont {L.~C.}\ \bibnamefont {McInnes}},\ and\
  \bibinfo {author} {\bibfnamefont {B.~F.}\ \bibnamefont {Smith}},\ }\bibfield
  {title} {\bibinfo {title} {Efficient management of parallelism in object
  oriented numerical software libraries},\ }in\ \href@noop {} {\emph {\bibinfo
  {booktitle} {Modern Software Tools in Scientific Computing}}},\ \bibinfo
  {editor} {edited by\ \bibinfo {editor} {\bibfnamefont {E.}~\bibnamefont
  {Arge}}, \bibinfo {editor} {\bibfnamefont {A.~M.}\ \bibnamefont {Bruaset}},\
  and\ \bibinfo {editor} {\bibfnamefont {H.~P.}\ \bibnamefont {Langtangen}}}\
  (\bibinfo  {publisher} {Birkh{\"{a}}user Press},\ \bibinfo {year} {1997})\
  pp.\ \bibinfo {pages} {163--202}\BibitemShut {NoStop}%
\bibitem [{\citenamefont {Choptuik}\ \emph {et~al.}(2003)\citenamefont
  {Choptuik}, \citenamefont {Lehner}, \citenamefont {Olabarrieta},
  \citenamefont {Petryk}, \citenamefont {Pretorius},\ and\ \citenamefont
  {Villegas}}]{Choptuik:2003qd}%
  \BibitemOpen
  \bibfield  {author} {\bibinfo {author} {\bibfnamefont {M.~W.}\ \bibnamefont
  {Choptuik}}, \bibinfo {author} {\bibfnamefont {L.}~\bibnamefont {Lehner}},
  \bibinfo {author} {\bibfnamefont {I.}~\bibnamefont {Olabarrieta}}, \bibinfo
  {author} {\bibfnamefont {R.}~\bibnamefont {Petryk}}, \bibinfo {author}
  {\bibfnamefont {F.}~\bibnamefont {Pretorius}},\ and\ \bibinfo {author}
  {\bibfnamefont {H.}~\bibnamefont {Villegas}},\ }\bibfield  {title} {\bibinfo
  {title} {{Towards the final fate of an unstable black string}},\ }\href
  {https://doi.org/10.1103/PhysRevD.68.044001} {\bibfield  {journal} {\bibinfo
  {journal} {Phys. Rev. D}\ }\textbf {\bibinfo {volume} {68}},\ \bibinfo
  {pages} {044001} (\bibinfo {year} {2003})},\ \Eprint
  {https://arxiv.org/abs/gr-qc/0304085} {arXiv:gr-qc/0304085} \BibitemShut
  {NoStop}%
\bibitem [{\citenamefont {Harmark}(2004)}]{Harmark:2003yz}%
  \BibitemOpen
  \bibfield  {author} {\bibinfo {author} {\bibfnamefont {T.}~\bibnamefont
  {Harmark}},\ }\bibfield  {title} {\bibinfo {title} {{Small black holes on
  cylinders}},\ }\href {https://doi.org/10.1103/PhysRevD.69.104015} {\bibfield
  {journal} {\bibinfo  {journal} {Phys. Rev. D}\ }\textbf {\bibinfo {volume}
  {69}},\ \bibinfo {pages} {104015} (\bibinfo {year} {2004})},\ \Eprint
  {https://arxiv.org/abs/hep-th/0310259} {arXiv:hep-th/0310259} \BibitemShut
  {NoStop}%
\bibitem [{\citenamefont {Dias}\ \emph {et~al.}(2007)\citenamefont {Dias},
  \citenamefont {Harmark}, \citenamefont {Myers},\ and\ \citenamefont
  {Obers}}]{Dias:2007hg}%
  \BibitemOpen
  \bibfield  {author} {\bibinfo {author} {\bibfnamefont {O.~J.~C.}\
  \bibnamefont {Dias}}, \bibinfo {author} {\bibfnamefont {T.}~\bibnamefont
  {Harmark}}, \bibinfo {author} {\bibfnamefont {R.~C.}\ \bibnamefont {Myers}},\
  and\ \bibinfo {author} {\bibfnamefont {N.~A.}\ \bibnamefont {Obers}},\
  }\bibfield  {title} {\bibinfo {title} {{Multi-black hole configurations on
  the cylinder}},\ }\href {https://doi.org/10.1103/PhysRevD.76.104025}
  {\bibfield  {journal} {\bibinfo  {journal} {Phys. Rev. D}\ }\textbf {\bibinfo
  {volume} {76}},\ \bibinfo {pages} {104025} (\bibinfo {year} {2007})},\
  \Eprint {https://arxiv.org/abs/0706.3645} {arXiv:0706.3645 [hep-th]}
  \BibitemShut {NoStop}%
\bibitem [{\citenamefont {Witek}\ \emph {et~al.}(2011)\citenamefont {Witek},
  \citenamefont {Cardoso}, \citenamefont {Gualtieri}, \citenamefont {Herdeiro},
  \citenamefont {Sperhake},\ and\ \citenamefont {Zilhao}}]{Witek:2010az}%
  \BibitemOpen
  \bibfield  {author} {\bibinfo {author} {\bibfnamefont {H.}~\bibnamefont
  {Witek}}, \bibinfo {author} {\bibfnamefont {V.}~\bibnamefont {Cardoso}},
  \bibinfo {author} {\bibfnamefont {L.}~\bibnamefont {Gualtieri}}, \bibinfo
  {author} {\bibfnamefont {C.}~\bibnamefont {Herdeiro}}, \bibinfo {author}
  {\bibfnamefont {U.}~\bibnamefont {Sperhake}},\ and\ \bibinfo {author}
  {\bibfnamefont {M.}~\bibnamefont {Zilhao}},\ }\bibfield  {title} {\bibinfo
  {title} {{Head-on collisions of unequal mass black holes in D=5
  dimensions}},\ }\href {https://doi.org/10.1103/PhysRevD.83.044017} {\bibfield
   {journal} {\bibinfo  {journal} {Phys. Rev. D}\ }\textbf {\bibinfo {volume}
  {83}},\ \bibinfo {pages} {044017} (\bibinfo {year} {2011})},\ \Eprint
  {https://arxiv.org/abs/1011.0742} {arXiv:1011.0742 [gr-qc]} \BibitemShut
  {NoStop}%
\bibitem [{\citenamefont {Emparan}(2020)}]{Emparan:2020vyf}%
  \BibitemOpen
  \bibfield  {author} {\bibinfo {author} {\bibfnamefont {R.}~\bibnamefont
  {Emparan}},\ }\bibfield  {title} {\bibinfo {title} {{Predictivity lost,
  predictivity regained: a Miltonian cosmic censorship conjecture}},\ }\href
  {https://doi.org/10.1142/S021827182043021X} {\bibfield  {journal} {\bibinfo
  {journal} {Int. J. Mod. Phys. D}\ }\textbf {\bibinfo {volume} {29}},\
  \bibinfo {pages} {2043021} (\bibinfo {year} {2020})},\ \Eprint
  {https://arxiv.org/abs/2005.07389} {arXiv:2005.07389 [hep-th]} \BibitemShut
  {NoStop}%
\bibitem [{\citenamefont {Anninos}\ \emph {et~al.}(1995)\citenamefont
  {Anninos}, \citenamefont {Bernstein}, \citenamefont {Brandt}, \citenamefont
  {Libson}, \citenamefont {Masso}, \citenamefont {Seidel}, \citenamefont
  {Smarr}, \citenamefont {Suen},\ and\ \citenamefont
  {Walker}}]{Anninos:1994ay}%
  \BibitemOpen
  \bibfield  {author} {\bibinfo {author} {\bibfnamefont {P.}~\bibnamefont
  {Anninos}}, \bibinfo {author} {\bibfnamefont {D.}~\bibnamefont {Bernstein}},
  \bibinfo {author} {\bibfnamefont {S.}~\bibnamefont {Brandt}}, \bibinfo
  {author} {\bibfnamefont {J.}~\bibnamefont {Libson}}, \bibinfo {author}
  {\bibfnamefont {J.}~\bibnamefont {Masso}}, \bibinfo {author} {\bibfnamefont
  {E.}~\bibnamefont {Seidel}}, \bibinfo {author} {\bibfnamefont
  {L.}~\bibnamefont {Smarr}}, \bibinfo {author} {\bibfnamefont {W.-M.}\
  \bibnamefont {Suen}},\ and\ \bibinfo {author} {\bibfnamefont
  {P.}~\bibnamefont {Walker}},\ }\bibfield  {title} {\bibinfo {title}
  {{Dynamics of apparent and event horizons}},\ }\href
  {https://doi.org/10.1103/PhysRevLett.74.630} {\bibfield  {journal} {\bibinfo
  {journal} {Phys. Rev. Lett.}\ }\textbf {\bibinfo {volume} {74}},\ \bibinfo
  {pages} {630} (\bibinfo {year} {1995})},\ \Eprint
  {https://arxiv.org/abs/gr-qc/9403011} {arXiv:gr-qc/9403011} \BibitemShut
  {NoStop}%
\bibitem [{\citenamefont {Libson}\ \emph {et~al.}(1996)\citenamefont {Libson},
  \citenamefont {Masso}, \citenamefont {Seidel}, \citenamefont {Suen},\ and\
  \citenamefont {Walker}}]{Libson:1994dk}%
  \BibitemOpen
  \bibfield  {author} {\bibinfo {author} {\bibfnamefont {J.}~\bibnamefont
  {Libson}}, \bibinfo {author} {\bibfnamefont {J.}~\bibnamefont {Masso}},
  \bibinfo {author} {\bibfnamefont {E.}~\bibnamefont {Seidel}}, \bibinfo
  {author} {\bibfnamefont {W.-M.}\ \bibnamefont {Suen}},\ and\ \bibinfo
  {author} {\bibfnamefont {P.}~\bibnamefont {Walker}},\ }\bibfield  {title}
  {\bibinfo {title} {{Event horizons in numerical relativity. 1: Methods and
  tests}},\ }\href {https://doi.org/10.1103/PhysRevD.53.4335} {\bibfield
  {journal} {\bibinfo  {journal} {Phys. Rev. D}\ }\textbf {\bibinfo {volume}
  {53}},\ \bibinfo {pages} {4335} (\bibinfo {year} {1996})},\ \Eprint
  {https://arxiv.org/abs/gr-qc/9412068} {arXiv:gr-qc/9412068} \BibitemShut
  {NoStop}%
\bibitem [{\citenamefont {Kov\'acs}\ and\ \citenamefont
  {Reall}(2020{\natexlab{a}})}]{Kovacs:2020pns}%
  \BibitemOpen
  \bibfield  {author} {\bibinfo {author} {\bibfnamefont {A.~D.}\ \bibnamefont
  {Kov\'acs}}\ and\ \bibinfo {author} {\bibfnamefont {H.~S.}\ \bibnamefont
  {Reall}},\ }\bibfield  {title} {\bibinfo {title} {{Well-Posed Formulation of
  Scalar-Tensor Effective Field Theory}},\ }\href
  {https://doi.org/10.1103/PhysRevLett.124.221101} {\bibfield  {journal}
  {\bibinfo  {journal} {Phys. Rev. Lett.}\ }\textbf {\bibinfo {volume} {124}},\
  \bibinfo {pages} {221101} (\bibinfo {year} {2020}{\natexlab{a}})},\ \Eprint
  {https://arxiv.org/abs/2003.04327} {arXiv:2003.04327 [gr-qc]} \BibitemShut
  {NoStop}%
\bibitem [{\citenamefont {Kov\'acs}\ and\ \citenamefont
  {Reall}(2020{\natexlab{b}})}]{Kovacs:2020ywu}%
  \BibitemOpen
  \bibfield  {author} {\bibinfo {author} {\bibfnamefont {A.~D.}\ \bibnamefont
  {Kov\'acs}}\ and\ \bibinfo {author} {\bibfnamefont {H.~S.}\ \bibnamefont
  {Reall}},\ }\bibfield  {title} {\bibinfo {title} {{Well-posed formulation of
  Lovelock and Horndeski theories}},\ }\href
  {https://doi.org/10.1103/PhysRevD.101.124003} {\bibfield  {journal} {\bibinfo
   {journal} {Phys. Rev. D}\ }\textbf {\bibinfo {volume} {101}},\ \bibinfo
  {pages} {124003} (\bibinfo {year} {2020}{\natexlab{b}})},\ \Eprint
  {https://arxiv.org/abs/2003.08398} {arXiv:2003.08398 [gr-qc]} \BibitemShut
  {NoStop}%
\bibitem [{\citenamefont {Arest\'e~Sal\'o}\ \emph {et~al.}(2022)\citenamefont
  {Arest\'e~Sal\'o}, \citenamefont {Clough},\ and\ \citenamefont
  {Figueras}}]{AresteSalo:2022hua}%
  \BibitemOpen
  \bibfield  {author} {\bibinfo {author} {\bibfnamefont {L.}~\bibnamefont
  {Arest\'e~Sal\'o}}, \bibinfo {author} {\bibfnamefont {K.}~\bibnamefont
  {Clough}},\ and\ \bibinfo {author} {\bibfnamefont {P.}~\bibnamefont
  {Figueras}},\ }\bibfield  {title} {\bibinfo {title} {{Well-posedness of the
  four-derivative scalar-tensor theory of gravity in singularity avoiding
  coordinates}},\ }\Eprint {https://arxiv.org/abs/2208.14470} {arXiv:2208.14470
  [gr-qc]}  (\bibinfo {year} {2022})\BibitemShut {NoStop}%
\bibitem [{\citenamefont {King}\ \emph {et~al.}(2017)\citenamefont {King},
  \citenamefont {Butcher},\ and\ \citenamefont {Zalewski}}]{apocrita}%
  \BibitemOpen
  \bibfield  {author} {\bibinfo {author} {\bibfnamefont {T.}~\bibnamefont
  {King}}, \bibinfo {author} {\bibfnamefont {S.}~\bibnamefont {Butcher}},\ and\
  \bibinfo {author} {\bibfnamefont {L.}~\bibnamefont {Zalewski}},\ }\href
  {https://doi.org/10.5281/zenodo.438045} {\emph {\bibinfo {title} {{Apocrita -
  High Performance Computing Cluster for Queen Mary University of London}}}}
  (\bibinfo {year} {2017})\BibitemShut {NoStop}%
\end{thebibliography}%

\end{document}